\documentclass[a4paper,11pt]{article}
\usepackage{jcappub} 
\usepackage{lineno}
\arxivnumber{2407.13385} 

\usepackage{mathtools}
\usepackage{graphicx}
\usepackage{subfig}
\usepackage{amsmath}
\usepackage{stix}
\usepackage{amssymb}
\usepackage{float}
\usepackage{xcolor}

\title{\boldmath The Stochastic Gravitational Wave Background from Primordial Gravitational Atoms}

\author{Zhaofeng Kang$^1$}
\affiliation{$^1$School of physics, Huazhong University of Science and Technology, Wuhan 430074, P. R. China}
\emailAdd{Zhaofengkang@gmail.com}

\author{Tianjun Li$^{2,3,4}$}
\affiliation{$^2$ CAS Key Laboratory of Theoretical Physics, Institute of Theoretical Physics, Chinese Academy of Sciences, Beijing 100190, P. R. China}
\affiliation{$^3$ School of Physical Sciences, University of Chinese Academy of Sciences, Beijing, P. R. China}
\affiliation{$^4$ School of Physics, Henan Normal University, Xinxiang, Henan 453007, P. R. China}
\emailAdd{tli@itp.ac.cn}
 
\author{and Weitao Ye$^1$}
\emailAdd{weitao.ye@hotmail.com}

\abstract{We propose a scenario of primordial gravitational atoms (PGAs), which may exist in the current and past universe due to spinning primordial black holes (PBHs) and very light bosonic fields. In a monochromatic mass scenario with a sizable dimensionless spin, which may arise in a short matter dominated (MD) era, we analyze the resulting stochastic gravitational wave background (SGWB) signal. Its spectrum is approximately characterized by a rising $\propto f^3$ followed by a falling $\propto f^{-1}$ where $f$ is the frequency. Then, we investigate the constraints and prospects of such an SGWB, and find that PGAs with a core mass $M_{\rm BH}\sim {\cal O}(10)~M_{\odot}$ and a cloud of light scalar with mass  $\mu \sim {\cal O} (10^{-13})$ eV could yield constraints even stronger than those from bare PBHs. Future detectors such as LISA, Taiji and TianQin are able to explore PGAs over a narrow and elongated strap in the $(\mu,M_{\rm BH})$ plane, spanning over 10 orders of magnitude for the maximum spin, $10^{-8}~M_{\odot}\lesssim M_{\rm BH}\lesssim 10^4~M_{\odot}$, $10^{-16}~{\rm eV}\lesssim \mu\lesssim 10^{-3}~\rm eV$.  If the PGA is dressed with a vector cloud, the SGWB signal has a much better opportunity to be probed.}

\begin{document}
\maketitle
\flushbottom

\section{Introduction}

The superradiance of rotating black holes (Kerr black holes) is a particularly intriguing phenomenon. Bosonic waves incident on a black hole satisfying the superradiance condition ($\omega<m\Omega_H$, where $\omega$ is the wave frequency, $m$ is the wave's angular frequency, and $\Omega_H$ is the black hole's angular velocity) will extract energy and angular momentum from the black hole, enhancing the reflected waves. Brito et al. provided a comprehensive review of the superradiance phenomenon in Ref. \cite{Brito_2020}. Based on it, Press and Teukolsky proposed the concept of superradiant instability \cite{press1972floating,cardoso2004black}. Imagine a scenario where the black hole is enveloped by a reflective mirror. Bosonic waves satisfying the superradiance condition would be trapped between the black hole and its mirror, continuously reflecting and thereby extracting energy from the black hole, enhancing the wave's energy. Considering practical scenarios, if there exist massive bosonic waves that satisfy the superradiance condition, gravity would act as a mirror, confining the bosonic waves near the black hole and forming quasi-normal bound states \cite{damour1976quantum,zouros1979instabilities,Detweiler:1980uk,furuhashi2004instability,cardoso2005superradiant,dolan2007instability,rosa2010extremal,dolan2013superradiant, east2017superradiant,east2017superradiant(2),dolan2018instability,bao2022imprzoved}. We refer to these bound states as "bosonic clouds" and designate the black holes enveloped by boson clouds as "gravitational atoms (GAs)". Bosonic clouds can radiate gravitational waves (GWs), providing a means to detect bosons, as reviewed in \cite{brito2020black}.

It is of great interest to consider that the core of the GA is a primordial black hole (PBH), which is believed to be relics of the early universe inhomogeneities~\cite{zel1967d,hawking1971gravitationally,carr1974black,chapline1975cosmological}~\footnote{Depending on their mass range, PBHs could potentially explain all or part of dark matter in the universe, as recently reviewed in \cite{carr2020primordial,green2021primordial}. The first detection of binary black hole mergers by LIGO/VIRGO in 2016 \cite{abbott2016observation} raised the possibility that the observed black holes might be PBHs~\cite{bird2016did,clesse2017clustering,sasaki2016primordial}, thus garnering increased attention for PBHs. 
The review \cite{carr2023observational} has summarized positive evidence for the existence of PBHs.
However, various studies have also constrained the maximum abundance of PBHs in different mass ranges, as in  \cite{carr2021constraints}. }. Therefore, primordial gravitational atoms (PGAs) convey information about the early universe encoded in the stochastic gravitational wave background (SGWB). One immediate piece of  information is on the spin of PBHs. 
In the standard cosmology,  PBHs only formed in the radiation-dominated (RD) era. The details of the standard scenario is discussed in Ref. \cite{harada2013threshold} and references therein.
It is widely believed that PBHs usually carry little spin, namely the dimensionless spin parameter $\tilde{a} \equiv \frac{M_{\rm PL}^2}{M_\text{BH}^2} J_\text{BH}\lesssim 0.01$, with $J_\text{BH}$ the angular momentum of the black hole~\cite{chiba2017spin,de2019initial,mirbabayi2020spin}, making it difficult for them to exhibit superradiance. 

The situation becomes different in non-standard scenarios, including both the non-standard origins of inhomogeneities other than the primordial quantum fluctuations during the inflation stage, as well as the non-standard phases of the early universe. Many scenarios of PBH formation are reviewed in Refs.~\cite{carr2020primordial,carr2021constraints}. In most scenarios, their spins are not clear, but there are still some hopeful cases which may give rise to a sizable $\tilde a$. For instance, it is shown that PBHs from collapsing domain walls may have $\tilde{a} \lesssim 0.1$  only for $M_{\rm BH} \gtrsim 10^{-3}M_\odot$, whereas less massive black holes receive extreme spins $\tilde{a} \simeq 1$~\cite{Eroshenko:2021sez}. A well-known example is of the collapse of over-density to PBH during the pressureless  matter-dominated (MD) era, where collapse is more prone to be anisotropic and their spins are typically large, even approaching 1 \cite{harada2017spins}.  But this study does not take into account matter accretion, which may play an important role in black hole formation and growth~\cite{DeLuca:2021pls}. Numerical relativistic studies focus on this topic~\cite{de2022primordial}. Finally, mergers of PBH may also result in a spinning remnant~\cite{barausse2009predicting}. Therefore, it makes sense to study the possible signals of superradiance associated with PGAs, which are distinguished from those associated with astronomical GAs. 

Actually, during the progress of this project, several works studying the same topic have already appeared, but they focus on different aspects of the PGAs. Pani and Loeb \cite{pani2013constraining} investigated PBHs surrounded by photon clouds. During the RD era, PBHs are enveloped by plasma, which imparts an effective mass to photons, resulting in the formation of PGAs. Considering the impact of photon clouds on the CMB, they placed constraints on the abundance of PBHs in the mass range of $10^{-8}M_\odot$ to $10^{0.2}M_\odot$. The constraints depend on the average initial spin of the PBHs, $\langle \tilde{a} \rangle$.
Ferraz et al. \cite{ferraz2022superradiant} investigated the electromagnetic signals produced by pion clouds around very light PBH of mass $M_\text{BH} \sim 10^{-18} M_\odot$. They found that efficient pion production is only possible for $\tilde{a} \gtrsim 0.8-0.9$, to overcome both neutral pion decay and charged pion annihilation. Branco et al.  \cite{branco2023superradiant} considered axion clouds with self-interactions, generated around asteroid-mass PBHs, which could constitute all dark matter. They found that a significant number of axions could escape from the bound states into the intergalactic medium or host galaxy, potentially contributing to the galactic and extragalactic background flux. By considering two specific cases with $\tilde{a}=0.01$ and $\tilde{a}=0.99$, they constrained the coupling constant of photons with $20~\text{eV}-10^6~\text{eV}$ axions. Recently, Dent et al. \cite{dent2024multi} also focused on these PGAs, conducting their research using multi-messenger probes. However, gravitational signal was not discussed until a very recent work~\cite{banerjee2024gravitational}, which considered the merger of two equal mass PBHs produced in first order phase transition, resulting in a PGA with  $\tilde{a}\sim0.67$.  The authors then studied the GWs of cloud transitions between different energy levels. 

Our focus is also the SGWB from PGAs, but our approach differs from Ref.\cite{banerjee2024gravitational}, which is motivated by the physics of first-order vacuum phase transitions. On the basis of it, they restrict the mechanism of PBH formation and spin generation. However, our goal is to explore the general features and potential of SGWB signals at detectors without restricting our discussion to specific generation mechanisms for PBHs or their spin values. We aim for a broader perspective with minimal assumptions. For comparison, we also consider a scenario where PBHs formed during a short MD era coexist with a real scalar or vector field in the universe. In this paper, we adopt a conservative approach to assume that the bosonic field only engages in gravity-related interactions and does not couple with other particle fields. Within this framework, we calculate the SGWB spectrum radiated by superradiant clouds surrounding the PBHs, to find that it is approximately characterized by a rising $\propto f^3$ followed by a falling $\propto f^{-1}$ where $f$ is the frequency. Using such a spectrum, the current data can place constraints on the abundance of these PBHs within the parameter space defined by the mass of the bosonic field $\mu$ and the mass of the PBHs $M_{\rm BH}$. The constraints can be even stronger than those made using bare PBHs. In future detectors such as LISA, Taiji and TianQin, in which a wide scalar cloud PGA population has the potential to leave imprints, roughly over $10^{-8}~M_{\odot}\lesssim M_{\rm BH}\lesssim 10^4~M_{\odot}$, $10^{-16}~{\rm eV}\lesssim \mu_S\lesssim 10^{-3}~\rm eV$. The spin of the vector boson is important, significantly increasing the discovery potential of PGAs with a vector cloud.

The work is organized as follows. In Section 2 we establish a simple scenario for PGAs. In Section 3 we calculate the SGWB spectrum from PGAs, and moreover analyze the current constraints on it and its prospects.  In the Appendix, we review the key details of PBHs formed in a,MD era. The final section includes the conclusions and discussions.
 
\section{PBH in the Monochromatic Mass Scenario } \label{sec short md}

The mass spectrum and spin distribution of PBHs, depend on their production mechanism, late accretion, and even their cosmic evolution~\cite{DeLuca:2020bjf}, so it is quite complicated to achieve a comprehensive understanding, which is beyond the scope of this work.  Although the actual mass spectrum of PBHs is extended, for specific cases such as primordial fluctuations peaking around a certain scale and the short MD era scenario considered next, the PBH mass can be approximated by a monochromatic spectrum. To simplify the problem, we first study the case in which the PBH mass follows a monochromatic distribution, which allows a tentative study of the GW signals from PGAs.  

In this paper, we consider a realistic scenario where PBHs are produced during a short MD era, and their spin distribution will be provided. Additionally, we will consider that the PBH mass and spin take fixed values $M_{\rm BH}$ and $\tilde{a}$, respectively, without targeting any specific formation scenario.

\subsection{Monochromatic PBHs from a Short MD Era }

Unlike the PBH formation scenario during the RD era, where pressure plays a major role, the formation of PBH during the MD era is primarily related to the anisotropy of the collapse. This has been studied in Ref. \cite{harada2016primordial}.  Furthermore, Harada  et al. \cite{harada2017spins} investigated the effects of the PBH spin on PBH formation, and also found that the nondimensional spins are typically large, even approaching 1. Ref. \cite{kokubu2018effect} examined the impact of inhomogeneity. Recently, Ref. \cite{harada2023threshold} studied the impacts of velocity dispersion. For numerical relativistic studies, see \cite{de2022primordial}. In order not to interfere with the main line of the article, we will include a review of the details in the Appendix~\ref{MD:PBH}, and in the following we just quote the key formulas.

In standard cosmology, the sole MD started at the time corresponding to matter-radiation equality. Since there is no observational data to constrain the thermal history of the universe before big bang nucleosynthesis (BBN) \cite{carroll2002testing,hooper2023dark}, various models of an early MD era before BBN exist.
These models include the domination of non-relativistic particles \cite{khlopov1980primordial,polnarev1982era,berlin2016pev,tenkanen2016reheating,allahverdi2022early}, the existence of moduli fields in string theory models \cite{green1997primordial,kane2015cosmological}, and the end of inflation \cite{albrecht1982reheating,kofman1994reheating,kofman1997towards,amin2015nonperturbative,carr2018primordial,aurrekoetxea2023oscillon,padilla2024primordial}.
We consider an early MD era lasts from $t_\text{ini}$ to $t_\text{end}$, during which the Hubble constant $H$ and scale factor $a$ respectively behave as
\begin{align}
    H=\frac{2}{a_m\eta^3}=2a_m^{1/2}a^{-3/2}=\frac{2}{3t} ~(t_\text{ini}<t<t_\text{end}),\quad a(t)=a_m \eta^2=a_m^{1/3}(3t)^{2/3},
\end{align}
with $a_m$ the integral constant determined by matching at $t_{\rm end}$ and $\eta$  the conformal time. Subsequently, these dominant matter would decay into high-energy particles as radiation, recovering the RD era. We assume that this happens instantly, and then the reheating temperature $T_\text{rh}$ is determined by energy conservation, $\rho(t_\text{end})=\rho_r (t_\text{end})$, giving $ T_\text{rh} = \left( {30 }/{\pi^2 g_*} \right)^{1/4} \rho(t_\text{end})^{1/4}$, where $g_{*}$ is the degree of freedom of radiation after $t=t_\text{end}$. After reheating, the universe undergoes an adiabatic expansion, and then both the energy density of PBH (behaving as matter) and the entropy density scale as $a^{-3}$, so the ratio $\rho_{\text{PBH}}(t>t_\text{end})/s(t>t_\text{end})$ keeps constant.  

In this paper, we consider PBHs formed from the collapse of primordial fluctuations after crossing the horizon, with masses approximately equal to the Hubble mass at the horizon crossing.
According to the discussion in Appendix  \ref{Appendix PBH mass}, a perturbation $\delta(k)$ entering the horizon at $t=t_H(k)$ will collapse at $t_c(k) \simeq t_H(k) \delta_H^{-3/2}(k) \simeq t_H(k) \sigma_H^{-3/2}(k)$, where $k$ denotes the comoving scale of the perturbation. It is manifest that for a successful collapse in the MD, it is required that $t_{\rm ini} < t_H(k) < t_c(k) < t_{\rm end}$.  For those perturbations with $t_c(k)>t_{\rm end}$, the universe enters the RD era before the collapse completes, significantly reducing the probability of PBH formation.
Thus, the shortest MD that can still form PBHs requires that perturbations entering the horizon at $t = t_{\rm ini}$ could collapse precisely at $t = t_{\rm end}$, leading to the condition:  
\begin{align}
    t_{\rm end} \simeq t_{\rm ini} \sigma^{-3/2}_{\rm max}
\end{align}
where $\sigma_{\rm max} \equiv \sigma_H(k_{\rm max})$, with $k_{\rm max}$ denoting the scale collapsing at $t=t_{\rm end}$. 
We refer to this MD scenario as the "short MD" scenario, but in fact, it may not be that short.
The PBH mass in the short MD scenario is $M_{\rm max}\simeq M_H(t_\text{end})\sigma_{\text{max}}^{3/2}$ as discussed in Appendix  \ref{Appendix PBH mass}, which implies that the mass spectrum is monochromatic and only the mode $k_{\rm max}$ can collapse into PBHs.

We arrange such an artificial setup to control the possible reduction of $\tilde a$ owing to matter accretion.  Recently, Ref.~\cite{de2023spinning} found that, the dimensionless spin $\tilde a $ reaches its peak after $\sim 1 $e-fold of collapse initiation. Then, $\tilde a$ suffers a fast reduction within  $\sim 1 $e-fold due to the accretion of non-rotating matter~\footnote{If this effect robust for a dust-like fluid rather than a classical field used in Ref.~\cite{de2023spinning} to mimic matter in the sense of state equation, still is an open question. For example, if the wave length of the particle becomes comparable or even shorter than that of the PBH horizon, the quantum effect becomes important and then the classical field analogy fails to capture that feature.}. A short MD era does not allow PBH to significantly accrete. Then, the corresponding PBH energy fraction approximately is given by~\cite{kokubu2018effect}
\begin{align}
\label{eq_1}
\frac{\rho_{\rm PBH}({t_{\rm end}})}{\rho({t_{\rm end}})}
= \beta(k_{\rm max})
\simeq 0.2055 \sigma_{\rm max}^{13/2},
\end{align}
where $\beta( k_{\rm max})$ represents the probability of perturbations at the scale \(k_{\max}\) collapsing into PBHs. For further details, please refer to Appendix \ref{MD:PBH}.

Otherwise, in the case with a long MD era, only the PBHs produced near the end of MD era is of our interest. But that requires a specific study on the constraints on the extended spectrum. Anyway, a more rigorous treatment for the general situation requires details of the primordial perturbation power spectrum and as well the beginning and end of the MD era. In the current work, we take the simplifying setup above and leave the latter for a future study. 
 
Moreover, for the setup we considered, the accretion effect can be ignored and then we assume that the spin distribution is given by~\cite{harada2017spins} . 
  \begin{gather}
  \label{spin_distribution}
    P_2(\tilde{a}) \propto \frac{1}{\tilde{a}^{5/3}} \exp 
		\left(
			- \frac{1}{2 \sigma_{\text{max}}^{2/3}} \left( \frac{2}{5} I\right)^{4/3} \frac{1}{\tilde{a}^{4/3}}
		\right),
    \end{gather}
which requires that the initial shapes of those overdense regions collapsing into PBHs are very close to spherical. $I$ is a dimensionless constant related to the perturbation power spectrum and we take $I=1$. For a detailed discussion on the production rate of PBHs and their spin distribution, see the Appendix \ref{MD:prsd}. Actually, the limit of monochromatic spectrum allows us to fully determine the above spin distribution of PBHs in terms of two observables, $M_{\rm BH}$ and $f_{\rm PBH}$, as shown below.

The key is to express the variance $\sigma_\text{max} $ in terms of $M_{\rm BH}$ and $f_{\rm PBH}$. To that end, we first use the radiation energy relation $\rho(t_{\rm end}) = 3s(t_{\rm end})T_{\rm rh}/4$ to get
 \begin{align}
 \label{eq2}
    \beta &= \frac{\rho_{\text{PBH}}(t_{\text{end}})}{\rho(t_{\text{end}})}
        = \frac{4}{3T_\text{rh}} \frac{\rho_{\text{PBH}}(t_{\text{end}})}{s(t_{\text{end}})}= \frac{4}{3T_\text{rh}} \frac{f_{\text{PBH}}\rho_\text{DM}}{s|_{ \rm present}},
 \end{align}
where $\rho_{\rm PBH}|_{\rm present}=\rho_\text{DM} f_\text{PBH}$, and $\rho_\text{DM}$ is the dark matter energy density at present. The reheating temperature can also be expressed in terms of the PBH mass and $\sigma_\text{max}$ as
\begin{align}
    T_\text{rh} &= \left( \cfrac{45}{4\pi^3 g_*} \right)^{1/4} M_{\rm PL}^{1/2} H(t_\text{end})^{1/2}  = \left( \cfrac{45}{4\pi^3 g_*} \right)^{1/4} M_{\rm PL}^{1/2} M_{\rm PL} (2M_\text{BH})^{-1/2} \sigma_\text{max}^{3/4},
\end{align}
where we have used the relation Eq.~\eqref{max:mass} to express $H(t_\text{end})$. Substituting it into Eq. \eqref{eq2}, we get a relation of the PBH formation probability and the PBH abundance
\begin{align}
        \label{eq_3}
		\beta = 2.6681 \times 10^{-9} \left( \frac{g_*} {106.75} \right)^{1/4} f_\text{PBH} \left( \frac{M_\text{BH}} {M_\odot} \right)^{1/2} \sigma_\text{max}^{-3/4}.
	\end{align}
Combining Eq. \eqref{eq_1} with Eq. \eqref{eq_3}, we eventually get the variance as desired
	\begin{align}
        \label{sigma_max}
		\sigma_\text{max} 
		&\simeq 0.0817 f_{\text{PBH}}^{4/29} \left( \frac{g_*} {106.75} \right)^{1/29} \left( \frac{M_\text{BH}}{M_\odot} \right)^{2/29}  \simeq 0.0817 f_\text{PBH}^{4/29} \left( \frac{M_\text{BH}}{M_\odot} \right)^{2/29}.
	\end{align}

In fact, the SGWB from PGAs in the short MD era scenario depends only on three parameters $M_{\rm BH}$, \(\mu\), and $f_{\rm PBH}$.

Later, we will need the birth time of PBHs $t_{\rm BH}$, which is approximated to be $t_\text{end}$, which is also the beginning of the RD era, and we can estimate it as follows. According to \cite{carr2021constraints}, we have
	\begin{align}
		\frac{M_{H}(t_{\rm end})}{M_\odot} 
		\simeq 2.03\times10^5 \left( \frac{t_{\rm end}}{1{\rm s}} \right)
		= 1.7539 \times 10^{10} \left( \frac{t_{\rm end}}{\text{1years}} \right).
	\end{align}
Therefore, we obtain
	\begin{align}
		t_\text{end} 
		&\simeq 5.7016\times10^{-11} \text{years} \frac{M_\text{BH}}{M_\odot} \sigma_\text{max}^{-3/2} \notag \\
		&\simeq 2.4418\times10^{-9} \text{years} \times \left( \frac{g_{*i}}{106.75} \right)^{-3/58} f_\text{PBH}^{-6/29} \left( \frac{M_\text{BH}}{M_\odot} \right)^{26/29}.
	\end{align}
Since the early MD era must end before BBN, therefore, PBHs must form before 1 second (e.g. $t_{\rm end}\lesssim 1{\rm s}$), which then imposes an upper bound on PBH mass, $M_\text{BH} \lesssim 10^5 M_\odot$.

\section{The Stochastic GW background from Primordial Gravitational Atoms}

The bosonic cloud, once it satisfies the superradiance condition, experiences exponential growth until it reaches saturation. During this growth, the energy and angular momentum of the GA's core are transferred to the bosonic cloud. After reaching saturation, the bosonic cloud starts to consume its own energy by emitting GWs. Over an extended period, the cloud eventually depletes. Since each PGA generates GWs, and given the randomness in the spin direction of PBHs and the phase of GWs, the GWs emitted by clouds overlap to form an isotropic and homogeneous SGWB in the universe. The observation of this SGWB can provide insight into the nature of PGAs. For simpilicity, we refer to the PGA clothed in a scalar cloud as PGA-0 and the PGA clothed in a vector cloud as PGA-1. 

In this section, we will first provide the necessary details of superradiance and GW radiation for two kinds of PGAs, and then derive and analyze the corresponding spectra of the SGWB, and finally investigate the signals on various current and planned GW detectors.

\subsection{Superradiance and GW Radiation: Scalar}

\subsubsection{The Superradiant Stage: Scalar Cloud Growth}

We consider the existence of a real scalar field in the universe, which is assumed to be a free field without interactions except for gravity.  We do not consider the effect of self-interactions of the scalar field, which may have a significant effect \cite{baryakhtar2021black}. Both aspects are model-dependent. 

The dynamics of the free real massive scalar field $\Phi$ in the Kerr spacetime is governed by the Klein-Gordon equation
\begin{align}
    (\nabla^\mu \nabla_\mu - \mu_S^2) \Phi = 0,
\end{align}
where $\nabla^\mu$ is the covariant derivative operator and $\mu_S$ is the mass of the scalar boson. In the Boyer-Lindquist coordinates, $x^\mu = [t,r,\theta,\phi]$, the solution can be solved by decomposing it into a linear combination of
\begin{align}
    \Phi_{\omega l m} = e^{-i\omega t} e^{im \phi} S_{lm}(\theta) R_{\omega l m} (r)+c.c.
\end{align}
By imposing appropriate boundary conditions, we can obtain solutions representing quasi-bound states.
These states can be labeled with $|nlm\rangle$, where $n= 1,2,3\ldots$, $l=0,1,\ldots n-1$, and $m=-l, -l+1,\ldots l$. Here $\omega$ is complex in general, with its real part $\omega_R$ representing the energy of the state, and its imaginary part $\omega_I$, representing the superradiant growth ($\omega_I > 0$) or the decay ($\omega_I < 0$); the quantum number $m$ represents the angular momentum of the state along the direction of the black hole spin. The gravitational fine-structure parameter, which is defined as $\alpha \equiv \mu_S M_\text{BH} / M_{\rm PL}^2 = 0.75 (M_\text{BH} / M_{\odot}) (\mu_S / 10^{-10}eV)$,  strongly influences the behavior of the solution. 

For $\alpha \ll 1$, the de Broglie wavelength of the scalar is larger than the gravitational radius of  the black hole. Since the black hole’s gravitational potential is Coulomb-like at large distances,  the energy level is obviously Hydrogen-like 
\begin{align}
    \omega_{R,nlm} = \mu_S \left(1 - \frac{\alpha^2}{2 n^2}\right)
    \label{hy} .
\end{align}
To obtain the analytical solution, the matched-asymptotic calculation is typically employed \cite{Detweiler:1980uk,baumann2019spectra}.
In this paper, we adopt the approximate result from \cite{baumann2019spectra}, 
\begin{align}
    \omega_{I,nlm} = \tilde{r}_+ C_{nl} G_{lm}(m\Omega_H - \omega_R) \alpha^{4l+5},
    \label{wi}
\end{align}
to leading order in $\alpha$ and $\tilde{a}$, where
\begin{gather}
    C_{nl} = \frac{2^{4l+1}(n+l)!}{n^{2l+4}(n-l-1)!}
    \left[ \frac{l!}{(2l)!(2l+1)!} \right]^2, \\
    G_{lm} = \prod_{k=1}^{l} \left[ k^2(1-\tilde{a}^2) + (\tilde{a}m-2\tilde{r}_+ \alpha \cdot \omega_R/\mu_S)^2 \right].
    \label{glm}
\end{gather}
In particular, $C_{21}=1/48$. The angular velocity of the black hole horizon
\begin{gather}
    \Omega_H = M_{\rm PL}^2\frac{\tilde{a}}{2M_\text{BH}\tilde{r}_+},
\end{gather}
with $\tilde{r}_+ \equiv 1+\sqrt{1 - \tilde{a}^2}$. Because the total energy of the scalar cloud $|nlm\rangle$, $M_C$ , is a bilinear function of $\Phi$, it is not difficult to obtain its growth
\begin{align}
	\frac{dM_C}{dt}=\Gamma_{nlm} M_C,
\end{align}
where $\Gamma_{nlm} = 2 \omega_{I,nlm}$, named as the "superradiant rate". As seen in Eq.~(\ref{wi}), the state with $\omega_R < m\Omega_H$ (this condition does not require $\alpha \ll 1$ in fact) is superradiant and will extract enengy and angular momentum from the black hole; the superradiant condition can be written in the form as
\begin{align}
	\tilde{a} > \frac{4 \alpha/m}{4 \alpha^2/m^2 + 1} \ 
	\& \& \
	\alpha < \frac{m}{2}.
\end{align}
where we have used $\omega_R \simeq \mu_S$. Immediately, the ground state with $l=m=0$ does not allow superradiance. Moreover, the factor $\alpha^{2l+5}$ in Eq.~(\ref{wi}) implies that the superradiant rates for high-$l$ modes are suppressed, so $l=m=1,n=l+1=2$ should be the fastest growing mode. 
As for $\alpha \gg 1$, by means of the WKB approximation, it is found that the mode with the fastest superradiant instability has \cite{zouros1979instabilities}
\begin{align}
    \omega_I = 10^{-7} \frac{M_{\rm PL}^2}{M_{\rm BH}} \exp(-1.84 \alpha),
\end{align}
which represents the instability is exponentially suppressed. Therefore, the large $\alpha$ case is not of interest in our work. 

Based on the numerical results in Ref. \cite{dolan2007instability} and the review in Ref. \cite{Brito_2020}, the fastest superradiant growth is observed for $|211\rangle$ at $\alpha \approx 0.42$ and $\tilde{a}\rightarrow 1$. Moreover, the superradiant rate of $|211\rangle$ is significantly higher than that of other modes, unless the $|211\rangle$ state does not satisfy the superradiance condition or is very close to the superradiance condition. Hence, in this paper, we focus solely on the specific scalar cloud, which imposes the lower and upper bound on the dimensionless spin and the gravitational fine constant, respectively
	\begin{align}
		\tilde{a} >\tilde{a}_s(\alpha)\equiv \frac{4 \alpha}{4 \alpha^2 + 1} \ 
		\& \& \
		\alpha < \frac{1}{2}
        \label{sr}.
	\end{align}
Since in our consideration the core mass remains constant (see blow), the above superradiance conditions will yield strong constraints in the $(\mu_S, M_{BH},\tilde a)$ parameter space. 

The evolution of the GA, composed of the black hole as the core and the cloud condensed from the scalar particles occupied the $|211\rangle$ state, satisfies the following equations which describe the energy and angular transfer between the core and cloud,
\begin{gather}\label{MCGROW}
	\frac{dM_C}{dt} = -\frac{dM_\text{BH}}{dt} = \Gamma_{211} M_C, \\
    \frac{dJ_C}{dt} = -\frac{dJ_\text{BH}}{dt} = \frac{1}{\mu_S}\frac{dM_C}{dt}.
\end{gather}
The equation in the last line utilizes the relation $J_C = N_C = M_C/\mu_S$, as the individual energy and angular momentum of the $N_C$  scalar bosons in the condensation cloud are $\mu_S$ and 1, respectively. Such a relation implies that the above two equations are not independent. Eq.~(\ref{MCGROW}) implies that the superradiance process initially exhibits exponential growth, and defines the characteristic timescale of cloud growth,
\begin{gather}\label{MCGROW:t}
	\tau_{\rm grow}\equiv 1/\Gamma_{211}=1/2\omega_{I,211}.
\end{gather}
We calculate $2\omega_{I,211}$ by the analytic approximate result in Eq.~(\ref{wi}), which can be well estimated by the initial mass and spin of the black hole. Although the results are strictly valid for $\alpha \ll 1$, we employ them for approximation purposes, as we only have $\alpha < 1/2$. 

If the superradiance condition is satisfied, then the cloud will grow exponentially until the
dimensionless spin of the black hole decreases to $\tilde{a}_s =  4 \alpha / (1 + 4 \alpha^2)$, at which the growth ceases. The maximum energy extracted from the black hole by the scalar cloud is about 10.78\%, as shown in \cite{hui2208black}. Therefore, we can neglect the variation of $M_{\rm BH}$, then a constant $\alpha$, which enables us to express the final energy of the scalar cloud as
	\begin{align}
        \label{cloud mass:max}
		M_C^\text{max}
		= \mu_S N_C^\text{max}
		= \mu_S J_C^\text{max}
		= \frac{\mu_S M_\text{BH}^2}{M_{\rm PL}^2} \left(\tilde{a}-\tilde{a}_s\right).
	\end{align}
The exponential growth of the cloud enables us to estimate the time for the cloud to reach saturation as $\tau_{\rm sat}=\ln(N_C^{\rm max})  \tau_{\rm grow}$. Due to the independence of $\tau_{\rm sat}$ from the initial occupation number of $|211\rangle$, even if it is very small (e.g. due to quantum fluctuations), a significant scalar cloud will eventually form. We will see that superradiance arises similarly for the spin-1 case. Therefore, the superradiance phenomenon provides us with an opportunity to study extremely light bosons. 

\subsubsection{The GW Emission Stage: Scalar Cloud Decay}
The story of the cloud is not over yet when it reaches saturation. For the real scalar field, the quasibound state satisfies $\Phi_{\omega l m} \propto \sin{\omega_R t}$, which means that the energy momentum tensor oscillates with an angular frequency of $2\omega_R \simeq 2\mu_S$, and therefore the cloud emits GWs with the same frequency \cite{siemonsen2020gravitational,yoshino2014gravitational}. From a microscopic perspective, the annihilation of two scalar particles at the same energy level produces a graviton with an energy of $2\omega_R$ \cite{arvanitaki2011exploring,arvanitaki2015discovering}. The power of GWs radiated from the scalar cloud leads to the decrease of the cloud energy $M_C$, and then one has
\begin{equation}
		\label{eq3}
		\frac{dM_C}{dt}= - M_{\rm PL}^2 \frac{d\tilde{E}}{dt} \frac{M_C^2}{M_\text{BH}^2},
	\end{equation}
where, in general, the rate $\frac{d\tilde{E}}{dt}$ is a function of $\alpha$ and $\tilde{a}$. But later we will see that, in our case of interest, the GW radiation matters merely after the scalar cloud saturates at $\tilde{a} =\tilde{a}_s(\alpha)$, so $\frac{d\tilde{E}}{dt}$ depends only on $\alpha$ by fixing $\tilde{a}$ at $\tilde{a}_s(\alpha)$.  For $|211\rangle$, its numerical expression is provided in Ref.~\cite{brito2017gravitational}. And recently, a good analytic approximation using quantum field theory is obtained \cite{yang2023gravitational}. 
\begin{align}
\label{dedt_scalar}
		\frac {d \tilde{E}} {dt}
		= &\frac{\alpha^{14}}{(2+\alpha^2) ^{11} (4+\alpha^2)^4} [
			6497.58 + 37469.8 \alpha^2 + 99032.5 \alpha^4 + 158627 \alpha^6  \notag\\
			&- \frac{4 \alpha^2 (1 - \alpha^2 / 8)}{1 + 4\alpha^2 (1 - \alpha^2 / 8)^2}
			(17329.2 + 95290.3 \alpha^2 + 240551 \alpha^4 + 368567\alpha^8)  \notag\\
			&+ \frac{16 \alpha^4 (1 - \alpha^2 / 8)^2}{(1 + 4\alpha^2 (1 - \alpha^2 / 8)^2)^2}	
			(11551.9 + 66235.9 \alpha^2 + 175871 \alpha^4 + 286598 \alpha^6) ].
	\end{align}
In this paper, for $\alpha > 0.1$, we use the numerical result, and for $\alpha < 0.1$, it is more convenient to adopt the above approximation. 
	
Let us show that it is justified to divide the evolution of the scalar cloud into two phases: the superradiant growth and the GW emission. If we ignore the change in $\alpha $, that is, considering that the PBH mass remains unchanged after scalar cloud saturation, then Eq.~\eqref{eq3} can be simply solved, giving 
	\begin{equation}
        \label{eq_mc}
		M_C(t) = \frac{M_C^\text{max}}{1+[t-t_0]/\tau_\text{GW}},
	\end{equation}
where $t_0$ is the cosmic time when the scalar cloud saturates, with mass $M_C^{\rm max}$ given in Eq.~\eqref{cloud mass:max}. Assuming the superradiance process initiates immediately after PBH formation, we have $t_0 \simeq t_{\rm birth} + \tau_{\rm sat}$ \footnote{It is worth noting that the formation time of PBHs, $t_{\rm birth}$, depends on the formation scenario of PBHs. For PBHs formed during a short MD era as we discussed, $t_{\rm birth} \simeq t_{\rm end}$. We also consider the even distribution of spin by fixing $\tilde a$, where we do not assume a specific formation scenario. Since PBHs form in the very early universe, $t_{\rm birth}$ is generally very small. Therefore, at this case, we neglect $t_{\rm birth}$, with $t_0 \simeq \tau_{\rm sat}$.}. The mass of the scalar cloud that decreases over time is characterized by the parameter 
	\begin{equation}
        \label{eq_taugw}
		\tau_\text{GW} = 1.6\times10^{-12}\text{years} 
				\left( \frac{d\tilde{E}}{dt} \right)^{-1}
				\left( \frac{M_\text{BH}}{M_\text{solar}} \right)^2
				\left( \frac{M_C^\text{max}}{M_\text{solar}} \right)^{-1},
	\end{equation}
which is the characteristic time for the consumption of the scalar cloud through the emission of GW. Now, we can compare the characteristic timescales for the exponential growth and GW emission:
	\begin{gather}
		\tau_\text{grow} \sim 7\times10^{-12}\text{years} \frac{M_\text{BH}}{M_\text{solar}} \alpha^{-9} \tilde{a}^{-1}, \\
		\tau_\text{GW} \sim 6\times10^{-12}\text{years} \frac{M_\text{BH}}{M_\text{solar}} \alpha^{-15} \tilde{a}^{-1}.
	\end{gather}
Although the above estimation is relatively rough,  
they are sufficient for a qualitative analysis. The superradiance of $|211\rangle$ requires $\alpha < 1/2$, therefore $\tau_{\rm GW}$ is significantly larger than $\tau_\text{grow}$.   
	
\subsection{Superradiance and GW Radiation: Vector}

The scenario of the superradiance phenomenon of a real massive vector field $ A^\mu$ is similar to scalar; the motion of $A^\mu$ satisfies the Proca equation
\begin{align}
    \nabla_\mu F^{\mu\nu} = \mu_V^2 A^\nu,
    \label{proca}
\end{align}
where $\mu_V$ is the mass of the vector boson and $F^{\mu\nu}=\partial^\mu A^\nu-\partial^\nu A^\mu$. From Eq.~(\ref{proca}), the Lorenz condition $\nabla_\mu A^\mu=0$ is obviously satisfied, which means the massive vector field has three intrinsic degrees of freedom. Different from the scalar, the quasi-bound states of the vector should be denoted by four quantum numbers $|nljm \rangle$ , where $l = 0, 1, ..., n-1$ are the quantum numbers for orbital angular momentum, $j = l-1, l, l+1$ represent the total angular momentum, and $m = 0, 1, ..., j$ represent the angular momentum along the direction of the black hole spin. 

When $\alpha \ll 1$ (unless specifically stated otherwise, the same symbols are defined as in the scalar, with only the physical quantities of the scalar field replaced by their corresponding ones for the vector field, such as replacing $\mu_S$ with $\mu_V$), the state energy is Hydrogen-like as shown in Eq.~(\ref{hy}), and the $\omega_I$ takes a very similar form \cite{baumann2019spectra} 
\begin{align}
    \omega_{I,nljm} = \tilde{r}_+ C_{nlj} G_{jm}(m\Omega_H - \omega_R) \alpha^{2l+2j+5},
    \label{}
\end{align}
where
\begin{gather}
    C_{nlj} = \frac{2^{2l+2j+1}(n+l)!}{n^{2l+4}(n-l-1)!}
    \left[ \frac{l!}{(l+j)!(l+j+1)!} \right]^2
    \left[ 1 + \frac{2(l-j+1)(j-l+1)}{l+j} \right]^2
\end{gather}
with $C_{101}=2$. $G_{jm}$ is given by replacing $l$ by $j$ in Eq.~(\ref{glm}). The fastest growing mode (refer to Ref.~\cite{east2017superradiant(2)} for numerical results) is the state $|1011\rangle$, whose superradiant rate $\Gamma_{1011}=2\omega_{I,1011}\propto \alpha^7$ is significantly larger than $\Gamma_{211}\propto \alpha^9$ in the scalar scenario for small $\alpha$. Similarly, in the vector scenario, we only consider this single mode.

For our consideration, the vector cloud composed of the $|1011\rangle$ particles, satisfying Eq.~(\ref{sr}) will continue growing until reaching saturation. Afterward, the GWs produced by the annihilation of vector particles will radiate away. The discussion is nearly identical to the previous section, except for the value of $\frac{d\tilde{E}}{dt}$, due to the enhanced GW emission from the vector cloud. It was analytically calculated to be $\frac{d\tilde{E}}{dt} = \frac{32}{5}\alpha^{10}$ for $\alpha\ll 1$ in the limit of flat spacetime~\cite{baryakhtar2017black}, and including corrections from a Schwarzschild geometry leads to an enhanced coefficient, 60. Later, numerical study improves the result to be~\cite{siemonsen2020gravitational} 
\begin{align}
\label{dedt_vector}
    \frac{d\tilde{E}}{dt} = 16.66\alpha^{10}, 
\end{align}
for $\alpha\ll 1$. For the same $\alpha$, it has a significantly higher GW emission rate $\frac{d\tilde{E}}{dt}$,  as seen in Eq. \eqref{dedt_scalar}. As a consequence, the cloud decay time $\tau_{\rm GW}\propto (\frac{d\tilde{E}}{dt})^{-1}$ for PGA-1 is much shorter compared to PGA-0 for small $\alpha$. This will be the main factor causing the observational difference between the two cases. 
		
\subsection{The SGWB Energy Density Frequency Spectrum}

PBHs are expected to be isotropically and homogeneously distributed throughout the universe, with their number per unit comoving volume given by
	\begin{equation}
		n=\frac{\rho_\text{DM}f_\text{PBH}}{M_\text{BH}},
	\end{equation} 
where PBHs follows a monochromatic mass distribution under our consideration.
The GWs emitted by the boson clouds have an initial frequency of $\mu / \pi$, but due to the cosmic expansion, this frequency is redshifted to $f = \frac{\mu}{ \pi (1+z)}$. Let us assume that the GW radiation lasts during the era where the standard $\Lambda$CDM model holds, and then the evolution of the red shift is given by
	\begin{gather}
		|\dot{z}| = (1+z)H = (1+z)H_0 \sqrt{\Omega_\Lambda + (1+z)^3 \Omega_m + (1+z)^4 \Omega_r}, 
	\end{gather}
where $H_0$ and $(\Omega_{\Lambda},\Omega_m,\Omega_r)$ represent the present-day Hubble constant and proportions of (dark energy, matter, radiation), respectively. Then, one can express the age of the universe with the observed frequency of GWs,
 \begin{gather}
		t(f) = \int_{\mu_S/(\pi f)}^\infty \frac{dz}{(1+z)H_0 \sqrt{\Omega_\Lambda + (1+z)^3 \Omega_m + (1+z)^4 \Omega_r}}.
	\end{gather}
Using Eq.~(\ref{eq3}), the energy density of GWs emitted from $t$ to $t+dt$ is
	\begin{equation}
		d\rho_{\rm GW}
		= -n \frac{d{M}_C}{dt} \frac{dt}{1+z} 
		= \frac{\rho_{\rm DM}f_{\rm PBH}}{M_{\rm BH}} M_{\rm PL}^2 \frac{d\tilde{E}}{dt} \frac{M_C^2}{M_{\rm BH}^2} \frac{df}{f|\dot{z}|},
	\end{equation}
with the factor $1/(1+z)$ accounting for the red shift of the frequency. To obtain the above expression, we have utilized the relation $dt/(1+z)=-df/f\dot{z}$. 

 Now, we calculate the energy density spectrum of the SGWB discussed earlier. Its general definition is
	\begin{equation}
		\Omega_{\rm GW}(f) \equiv \frac{f}{\rho_{c}} \frac{d \rho_{\rm GW}}{df},
	\end{equation}
where $\rho_{c}$ is the critical density at present. Therefore, the frequency spectrum of the SGWB energy density is given by
	\begin{equation}\label{spectrum}
		\Omega_{\rm GW}(f)= \frac{\Omega_{\rm DM} f_{\rm PBH}}{M_{\rm BH}} M_{\rm PL}^2 \frac{d\tilde{E}}{dt} \frac{M_C^2[t(f)]}{M_{\rm BH}^2 |\dot{z}|},
	\end{equation}
with $\Omega_{\rm DM}\equiv\rho_{\rm DM}/\rho_{c}$. Considering the distribution of spin and the superradiance condition, we ultimately obtain 
	\begin{equation}
        \label{eq_omega}
		\Omega_{\rm GW}(f) 
		= \theta \left( \frac{1}{2} - \alpha \right) 
		\frac{M_{\rm PL}^2}{M_{\rm BH}^3}
		\frac{\Omega_{\rm DM} f_{\rm PBH}}{|\dot{z}|} \frac{d\tilde{E}}{dt}
		\int^1_{4 \alpha/(4 \alpha^2 + 1)} P(\tilde{a}) d \tilde{a}\ M^2_C[t(f)],
	\end{equation}
which is determined by ($\mu,M_{\rm BH}$, $f_{\rm PBH}$) and the spin distribution of PBHs, $P(\tilde{a})$. In the following, we will consider the case where all PBHs have their spins fixed to the same value, as well as the scenario of the short MD era. The SGWB in the latter can be determined by only three parameters ($\mu,M_{\rm BH}$, $f_{\rm PBH}$) , because $P_2(\tilde{a})$ can be determined by $M_{\rm BH}$ and $f_{\rm PBH}$ as we discussed in the Section \ref{sec short md}.

It is evident that $f_{\rm PBH}$ determines the amplitude of the SGWB by directly determining the number of GW sources. In particular, in the short MD scenario, \( f_{\rm PBH} \) also indirectly influences the SGWB through the PBH spin distribution. However, given that the spin of PBHs is concentrated around magnitudes of order-1, this indirect effect is subdominant. Consequently, $\Omega_{\rm GW}(f) \propto f_{\rm PBH}$.

A significant distinction between PGAs and astronomical GAs is that the GWs of PGAs contain components from the early universe (i.e., the RD era). Therefore, it is instructive to derive the approximate behaviour of the SGWB spectrum generated during the RD era. For simplicity, we omit the influence of the spin distribution and initial time $t_0$. Note that in the RD era, the time $t$ corresponding to the frequency $f $ ($\propto a$) has $t(f)\propto a^2 \propto f^2$ and $|\dot{z}| \propto f^{-3}$.Writing $t=\tilde{\tau}(\frac{f}{\mu})^2$  with $\tilde{\tau}\equiv\pi^2t_{\rm eq}(1+z_{\rm eq})^2\sim 10^{12} \rm years$ , then, from Eq. \eqref{eq_mc} and Eq. \eqref{spectrum} we get the spectrum shape
\begin{align}
\Omega_{\rm GW}(f)\propto \frac{M_C^2}{|\dot{z}|} \propto \frac{f^3}{(1+\cfrac{\tilde{\tau}}{\tau_{\rm GW}}\cfrac{f^2}{\mu^2})^2}.
\end{align}
Other factors only determine the height of the spectrum, not its shape. The numerator, coming from $\dot z$, represents the effect of redshift on the GW spectrum. Hence, it means that the higher frequency GWs, which are produced at the nearer cosmological time and less affected by the redshift, have a relatively larger amplitude. The denominator reflects the effect of cloud decay over time, and conversely the higher $f$, produced at a time corresponding to a lighter cloud, leads to a reduction in the emitted GWs.

We can describe the above two competing effects more precisely, in terms of the characteristic cloud consumption time $\tau_{\rm GW}$. For the low frequency  $\frac{f}{\mu} \ll \sqrt{\frac{\tau_{\rm GW}}{\tilde{\tau}}}$ emitted at $t(f)\ll\tau_{\rm GW}$, the energy of the cloud remains nearly constant, thus we have the rising spectrum $\Omega(f) \propto f^3$; it is a power law widely present in SGWB signals due to the pure redshift. On the contrary, for the high frequency $\frac{f}{\mu}\gg \sqrt{\frac{\tau_{\rm GW}}{\tilde{\tau}}}$  produced at $t(f)\gg\tau_{\rm GW}$, the effect of the cloud decay becomes more important than the redshift, so we have the falling spectrum $\Omega(f) \propto f^{-1}$, which is the characteristic  power law specific to the SGWB from PGAs. More precisely, by solving $\frac{d\Omega_{\rm GW}(f)}{df}=0$, we find that the trend in the GW spectrum undergoes a transition at $f_{\rm peak}$ satisfying $\frac{f_{\rm peak}}{\mu} = \sqrt{\frac{3\tau_{\rm GW}}{\tilde{\tau}}}$ (i.e. $t(f_{\rm peak})=3\tau_{\rm GW})$, which is consistent with our physical expectations. 
In brief, the spectrum has an approximate behavior, as shown by

\begin{equation}
\label{omega_estimation}
    \Omega_{\rm GW} \propto
    \begin{cases}
        f^3 & \quad f \in (f_0, f_{\rm peak}) \\
    f_{\rm peak}^4 f^{-1} & \quad f \in (f_{\rm peak}, \frac{\mu}{\pi}).
    \end{cases}
\end{equation}

\begin{figure}[H]
    \centering
    \includegraphics[width=1\textwidth]{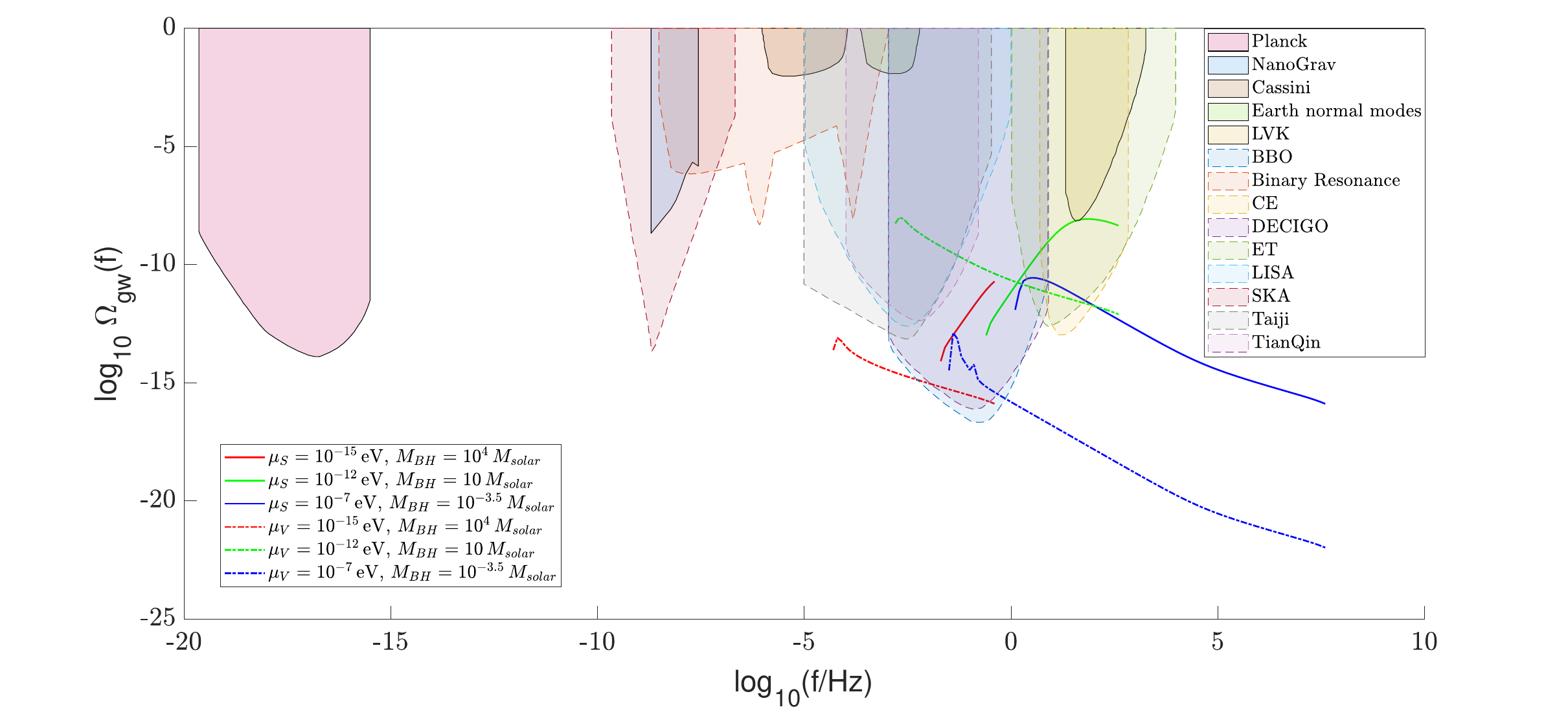} 
    \caption{Constraints on SGWBs from existing GW detectors (shaded areas with thin solid bound lines), sensitivities of future detectors (shaded areas with thin dashed bound lines), and the SGWBs from the PGAs in the short MD era scenario (solid lines for PGA-0 and dash-dotted lines for PGA-1). In our calculations, we choose $f_{\rm PBH}$ as the maximum value after considering all constraints, including those obtained from SGWBs in this study and other existing constraints summarized in Ref.~\cite{carr2021constraints}.}
    \label{GW_1}
\end{figure}

\begin{figure}[H]
  \centering            
  \subfloat[] 
  {
      \label{GW_2a}\includegraphics[width=0.49\textwidth]{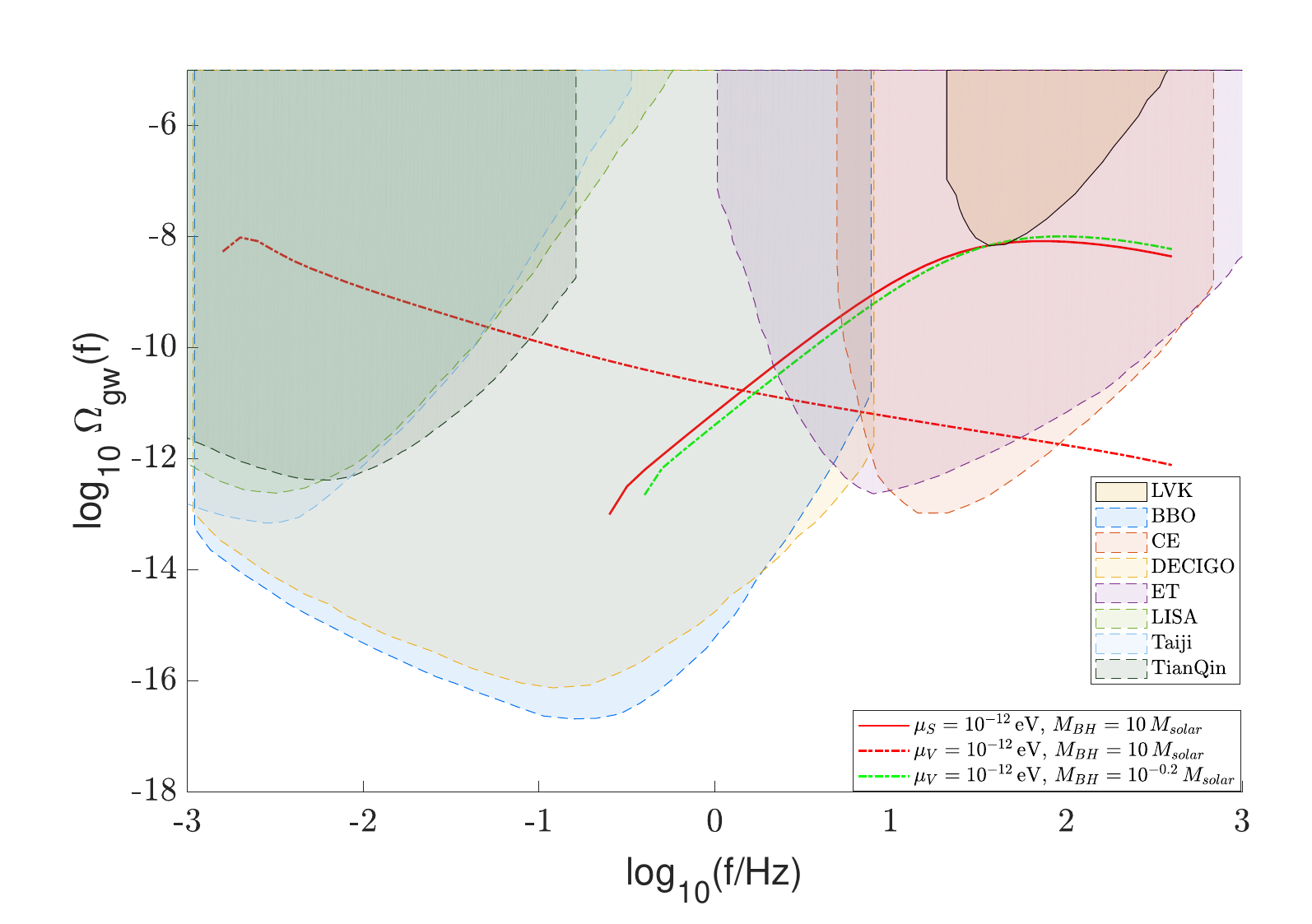}   
  }
  \subfloat[]
  {
      \label{GW_2b}\includegraphics[width=0.51\textwidth]{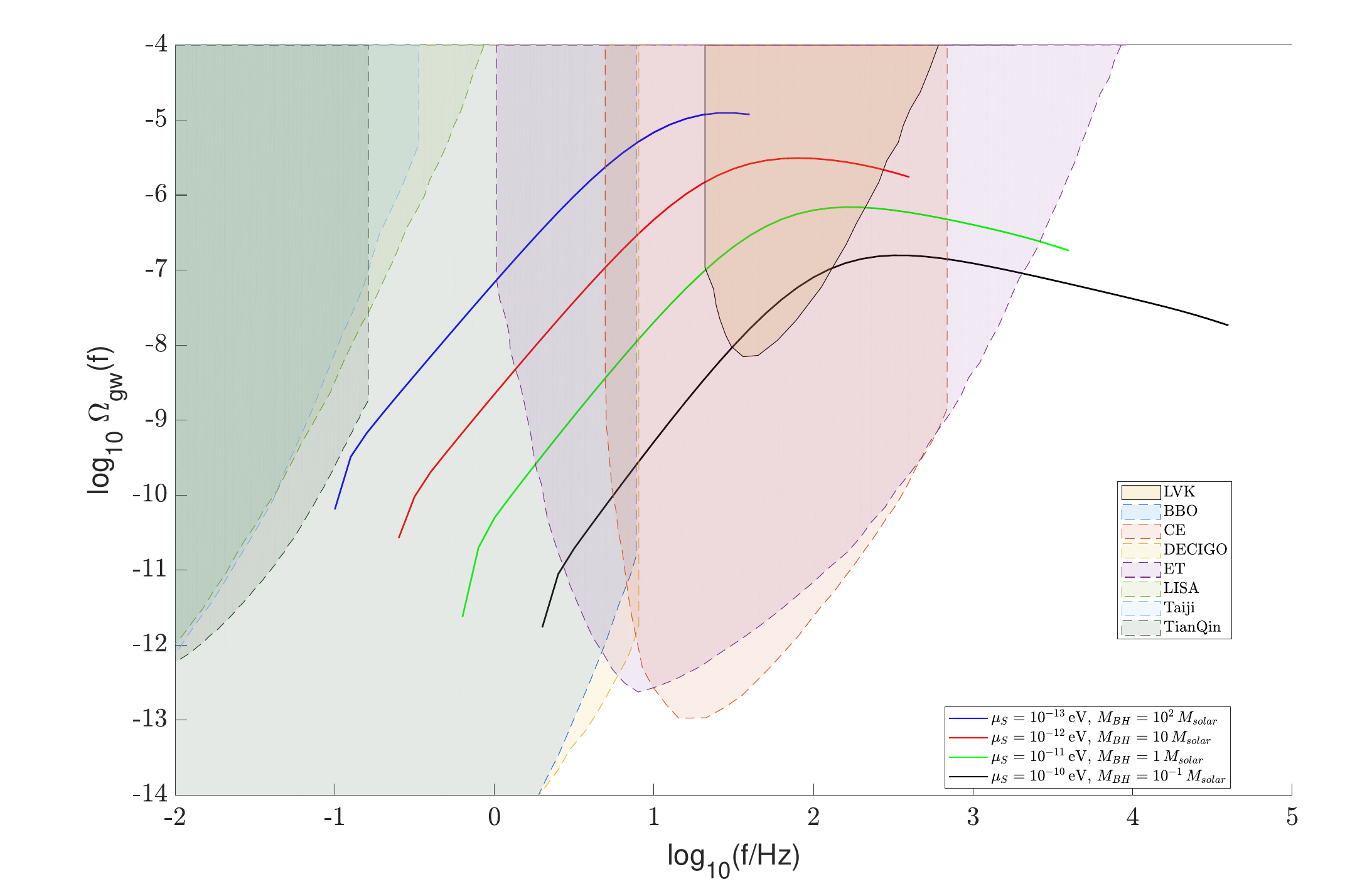}
  }
  \caption{Constraints on SGWBs from existing GW detectors (shaded areas with thin solid bound lines), sensitivities of future detectors (shaded areas with thin dashed bound lines), and the SGWBs from the PGAs in the short MD era scenario (solid lines for PGA-0 and dash-dotted lines for PGA-1). In our calculations, we choose $f_{\rm PBH}$ as the maximum value after considering all constraints, including those obtained from SGWBs in this study and other existing constraints summarized in Ref.~\cite{carr2021constraints}. (a) We select $f_{\rm PBH}$ as the maximum value after considering all constraints, including those obtained from SGWBs in this study and other existing constraints; (b) To reflect that the spectra produced by PGA with the same $\alpha$ vary monotonically, we set $f_{\rm PBH} = 10^{-3}$ .}  
  \label{GW_2}         
\end{figure}

To our knowledge, this new spectrum was obtained for the first time, and it is distinguished from other SGWB signals. Although the approximate spectrum is suitable only for GWs from the RD era, it is sufficient for a qualitative analysis of the complete spectrum, as shown in Fig. \ref{GW_1} and Fig. \ref{GW_2}.
According to $f = \frac{\mu}{ \pi (1+z)}$, the spectrum has two cuts. The lower cut $f_0$, originates from GWs emitted at time $t_0$, while the higher cut $\frac{\mu}{\pi}$ comes from GWs emitted today. 
In particular, for those GAs that emit GWs particularly slowly (where $\tau_{\rm GW}$ exceeds the age of the universe), $f_{\rm peak}>\frac{\mu}{\pi}$. This results in the absence of a peak frequency in their spectrum, leading to a spectrum that exhibits only a rising profile. The solid red line in Fig. \ref{GW_1} corresponds to this case.

To facilitate analyzing the SGWB at the GW detectors in the next subsection, it is helpful to elaborate on the behaviors of the spectra in the parameter space. They are derived from the fact that, in terms of Eq.~(\ref{omega_estimation}), the SGWB spectrum is determined mainly by the location $f_{\rm peak}$, which in turn depends on $\mu$ and is very sensitive to $\alpha$ since $f_{\rm peak} = \mu_S \sqrt{\tau_{\rm GW}}\propto \sqrt{\frac{\mu_S}{\tilde{a}-\tilde{a}_s}} \alpha^{-9}$ for PGA-0 and $f_{\rm peak} \propto \sqrt{\frac{\mu_V}{\tilde{a}-\tilde{a}_s}} \alpha^{-7}$ for PGA-1 (by the way, the coefficient of PGA-1 is smaller, which can be derived by comparing Eqs. (\ref{dedt_scalar}) and (\ref{dedt_vector})). The obvious difference between the red dash-dotted line and the green dash-dotted line in Fig. \ref{GW_2a} manifests this; they denote the spectra of PGA-1 having the same $\mu_V$ but different $\alpha = 0.075$ and $0.005$, respectively. On the other hand, for PGA-0 with the same $\alpha=0.075$ but increasing $\mu_S$ from left to right, the resulting spectra have similar shapes and almost shift monotonically towards the higher frequency region, as shown in Fig. \ref{GW_2b}.

Moreover, due to the different power law dependencies of PGA-0 and PGA-1 on $\alpha$ and the different coefficients, the peak frequency of PGA-1 is significantly lower than that of PGA-0 for the same $\mu$ and $\alpha$.
We show this by comparing the difference between the solid and dash-dotted lines in Fig. \ref{GW_1}. However, if we choose a smaller $\alpha$ for PGA-1 such that its peak frequency matches that of PGA-0, their spectra will become very similar, as shown by comparing the solid red line with $\alpha=0.075$ and the green dash-dotted line with $\alpha=0.005$ in Fig. \ref{GW_2a}.

\subsection{Results of SGWB from PGA: Constraints \& Prospects}

First of all, PBHs face a bunch of constraints, depending on the PBH mass, and they are comprehensively summarized in Ref.~\cite{carr2021constraints}. We assume that the dressed bosonic clouds do not affect these constraints, and explore PGAs in the allowed $(M_{\rm BH}, f_{\rm PBH})$ plane. Our interest is the SGWB signal from PGAs, and in this part we demonstrate the current constraints and explore its prospects at the future detectors. Despite of the overall similarity, we discuss the spin-0 and spin-1 cases separately.

\subsubsection{PGAs with a Scalar Cloud}

Let us first investigate the typical PGAs that can produce observational SGWBs in the monochromatic mass scenario. For demonstration, we consider the cores produced in the short MD era studied previously, taking $I=1$. The PGAs in the universe can be characterized by three parameters $(\mu_S, M_{\rm BH},f_{\rm PBH})$. Then, for each point of the parameter $(\mu_S,M_{\rm BH})$, we input different values $f_{\rm PBH}$ and calculate the GW spectrum from the corresponding PGAs. An upper bound $f^{\rm max}_{\rm PBH}(M_{\rm BH})$ is obtained by the requirement that the corresponding critical spectrum does not pass through any of the shaded areas with thin solid boundary lines shown in Fig.~\ref{GW_1}. These exclusions are derived from various available experiments that are sensitive to SGWBs, including the Planck CMB temperature and polarization spectra~\cite{lasky2016gravitational,akrami2020planck}, NanoGrav's approximate upper limit~\cite{agazie2023nanograv,NANOGrav:2023hvm}, Cassini satellite Doppler tracking~\cite{armstrong2003stochastic}, Earth's normal modes monitoring~\cite{coughlin2014constraining}, and Advanced LIGO's and Advanced Virgo's O3 run combined with limits from earlier O1 and O2 runs (LVK)~\cite{abbott2021upper}.  However, it is not difficult to observe that the constraints on the abundance of PBHs are entirely due to the contribution of LVK, as shown in Fig. \ref{GW_1}.

\begin{figure}[htbp]
			\centering
			\includegraphics[width=0.9\textwidth]{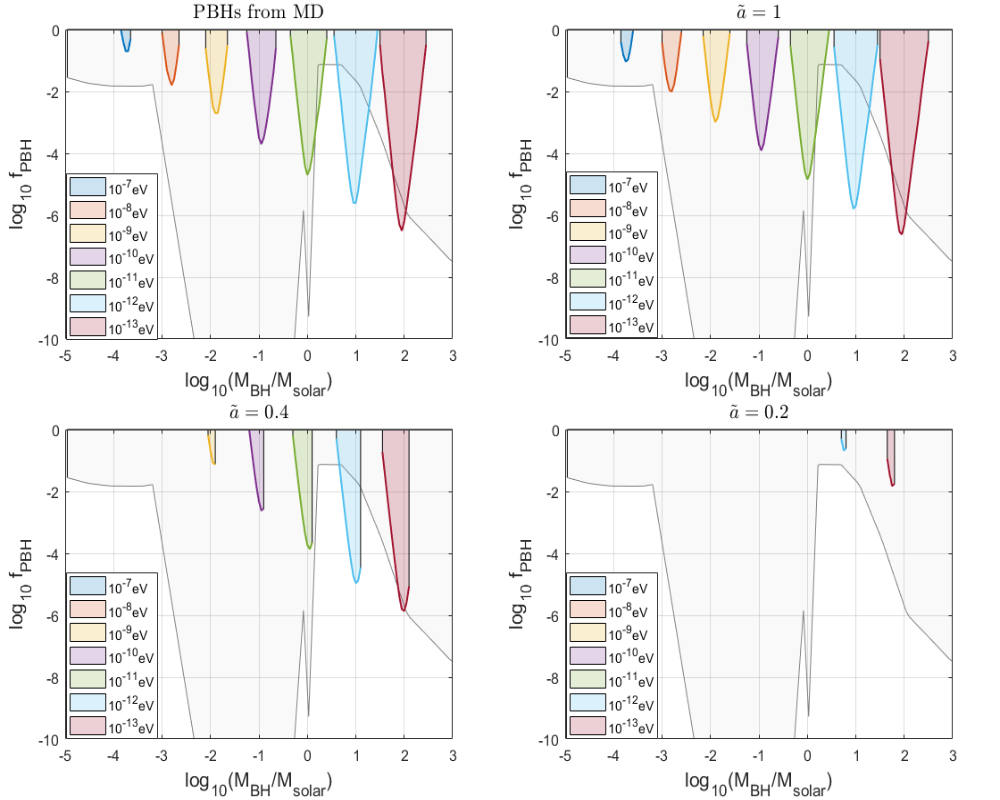}
			\caption{New constraint on PBHs by virtue of the SGWB signals from PGA-0 with several chosen $\mu_S$. The gray shaded areas denote the existing constraints. }
   \label{newconstraint}
		\end{figure}

In fact, in some PBH mass region, a proper cloud may increase the discovery of these PBHs via the SGWB signal. To show this, we add some plots in Fig.~\ref{newconstraint} where the strongest constraints are shaded in gray. It is seen that for $M_{\rm BH}\sim{\cal O}(1-100)~M_{\odot}$ subject to existing relatively weak constraints, the SGWB signal is already able to produce stronger constraints provided the scalar clouds with $\mu_S\sim{\cal O}(10^{-11}-10^{-13})~{\rm eV}$, which gives rise to GW spectra similar to the green solid line in Fig.~\ref{GW_1}; the constraints it provides even have several orders of magnitude enhancements.

Moreover, observing the resulting constraints on $f_{\rm PBH}$ in Fig.~\ref{newconstraint}, we find that they have the following characteristics:
\begin{itemize}
    \item The shapes of the constraint regions are similar and exhibit a "periodic translation." 
Taking $\mu_S$ = ($10^{-13}$, $10^{-12}$, $10^{-11}$, $10^{-10}$) eV as examples, the lowest points of these constraint regions occur at $M_{\rm BH}$ = ($10^{2}$, $10^{1}$, $10^{0}$, $10^{-1}$) $M_\odot$, all corresponding to the same $\alpha$. Clearly, this is not a coincidence. As mentioned at the end of the previous subsection, the spectrum of SGWB strongly depends on $\alpha$. For the same $\alpha$, the spectra are similar, and therefore the constraints on $f_{\rm PBH}$ should also be similar. Thus, the parameters corresponding to the minima of these constraint regions should have the same $\alpha$, which explains the similarities and periodicity observed in the constraint regions.
\item The strength of the constraint regions increases monotonically with $\mu_S$. In Fig. \ref{GW_2b}, we have presented the spectra of the four lowest points mentioned above, where we set $f_{\rm PBH}=10^{-3}$ for ease of comparison. It is clear that their spectra strengthen monotonically with increasing $\mu_S$. Consequently, this leads to a corresponding monotonic change in the constraint regions.
\end{itemize}

\begin{figure}[htbp]
			\centering
			\includegraphics[width=0.9\textwidth]{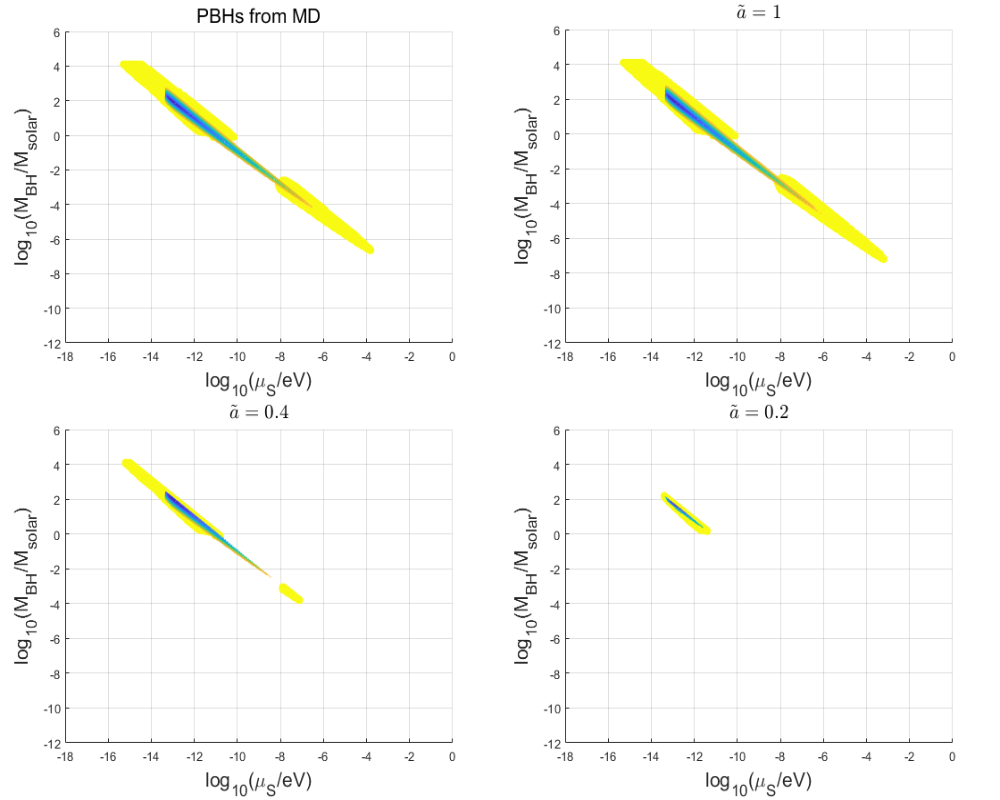} 
			\caption{Constraints (blue and green bands) and prospects (yellow bands) of PGA-0,  taking different spins $\tilde a$. }
       \label{paraspace_scalar}
		\end{figure}

To reveal the dependence of the spin, which is the peculiar feature of the PBHs that we are using and thus furnishes a way to distinguish the different origins of PBHs, we also scan the constraints on the $(\mu_S,M_{\rm BH})$ plane by choosing three values of fixed $\tilde a$: 1, 0.4 and 0.2. The results are displayed in Fig.~\ref{paraspace_scalar}, where the narrow green/blue shaded bands can be excluded. The area of the excluded band shrinks as $\tilde{a}$ decreases, and, in fact, we find that for $\tilde{a} = 0.1$, there are no constraints at all.
 
The future GW detectors in different frequency bands are about to substantially enhance the current sensitivities, and thus it is of interest to see the prospects of various PGAs at these future detectors. This analysis includes considerations of the sensitivities of various future GW detectors, such as the Square Kilometer Array (SKA) \cite{janssen2014gravitational}, binary resonance searches \cite{blas2022detecting,blas2022bridging}, Laser Interferometer Space Antenna (LISA) \cite{amaro2017laser}, Taiji \cite{ruan2020taiji}, TianQin\cite{liang2022science}, Deci-Hertz Interferometer Gravitational wave Observatory (DECIGO) \cite{kawamura2011japanese}, Big Bang Observer (BBO) \cite{phinney2004big}, Einstein Telescope (ET) \cite{punturo2010einstein}, and Cosmic Explorer (CE) \cite{reitze2019cosmic}. Their expected sensitivity curves are shown by the shaded areas with thin dashed bound lines in Fig.~\ref{GW_1}, where, we exhibit a group of promising $(M_{\rm BH},\mu)$ samples in the short MD era scenario, taking the upper limit of $f_{\rm PBH}(M_{\rm BH})$. One can see that the infrared part and the peak frequency of a given spectrum may be probed by different detectors, which have the best sensitivity at quite different bands. 

It is of importance to make a more general estimate on the GW prospects of PGAs. 
To that end, we consider the short MD era scenario and three fixed $\tilde{a}$ values: 1, 0.4, and 0.2. The regions of expected detectable parameters for these future detectors are shown in Fig. \ref{paraspace_scalar}. The yellow regions in the figure indicate the parameter spaces where the SGWB are expected to be detectable. In the mass window $M_{\rm BH}= 10^{-3}M_{\odot} \sim 1M_{\odot}$, no observable GWs are expected, due to the existing strict constraints on the abundance of PBHs in this mass range. 
If future observations do not detect signals in the yellow parameter spaces, it would imply even more stringent constraints on the abundance of PBHs in those regions.

\subsubsection{PGAs with a Vector Cloud}

Let us move on to the analysis of PGA-1. The results are found to be quite similar to those for PGA-0, except that the vector-boson spin brings a mild difference. We consider the short MD era scenario and three fixed values of $\tilde{a}$: 1, 0.1, and 0.04 for our analysis. We primarily discuss several notable differences between the detection of PGA-0 and PGA-1. 
\begin{itemize}
    \item Firstly, it is observed that for the same boson mass $\mu_V=\mu_S$, the PGA-1 mass at the lowest point of the constraint region shifts approximately $1~M_\odot$ to the left, compared to PGA-0, as illustrated in Fig. \ref{newconstraint} and Fig. \ref{newconstraint_vector}. To see how it happened, let us look at Fig. \ref{GW_2a}.
    We consider the sample $(\mu_S=\mu_V=10^{-12}~\rm eV, M_{BH}=10 ~M_{\odot})$. For PGA-0, the resulting SGWB near $f_{\rm peak}$ is constrained by LVK. However, for PGA-1, the resulting amplitude $\Omega_{\rm GW}(f)$ is significantly lower than the detection sensitivity of LVK due to the lower $f_{\rm peak} $ compared to PGA-0. So, for $\mu_V=\mu_S$, only the lighter PGA-1 (that is, smaller $\alpha$) can achieve a relatively higher $\tau_{\rm GW}$, allowing the SGWB to reach the LVK sensitivity.
\begin{figure}[htbp]
			\centering
			\includegraphics[width=0.9\textwidth]{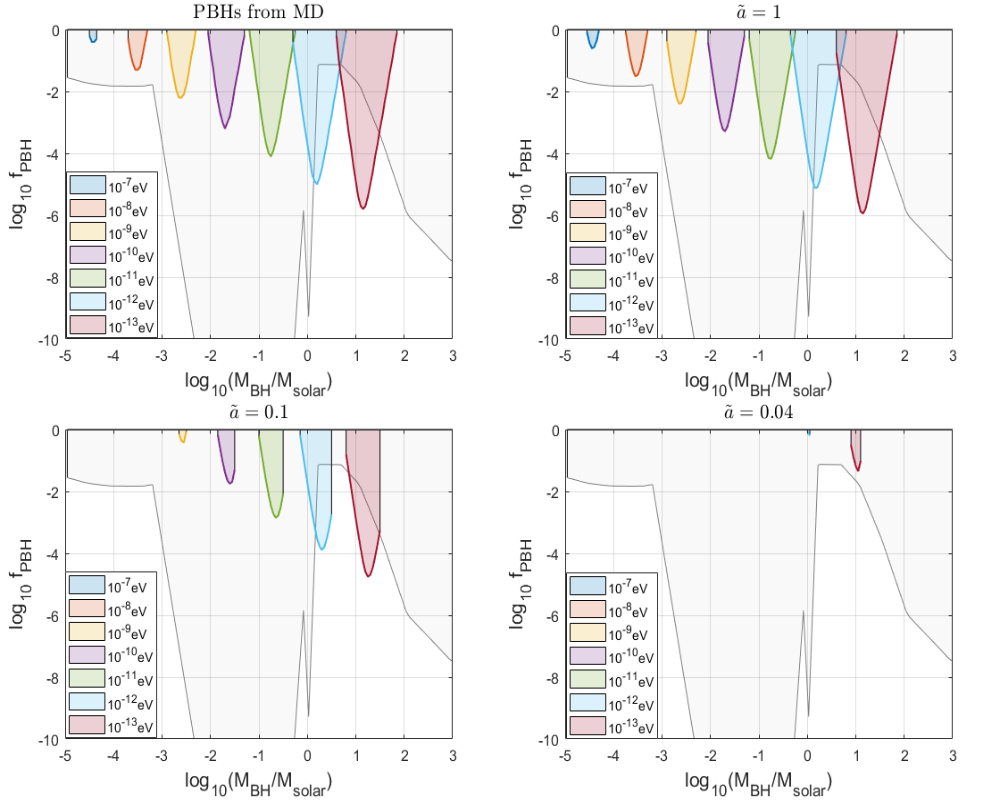} %
			\caption{New constraints on PBHs by virtue of the SGWB signals from PGA-1 with several chosen $\mu_V$. The gray-shaded areas denote the existing constraints. }
                \label{newconstraint_vector}
		\end{figure}
    
    \item Secondly,  PGA-1 still have a significant parameter space constrained by the SGWB for its core dimensionless spin values as low as \(\sim \mathcal{O}(0.01)\), which is obviously lower than the threshold for PGA-0 discussed earlier, \(\sim 0.1\) (compare Fig. \ref{newconstraint} with Fig. \ref{newconstraint_vector}). This is because the superradiance condition in Eq.~\eqref{sr} selects certain PGAs with \(\tilde{a} > \tilde{a}_s \equiv \frac{4 \alpha}{4 \alpha^2 + 1}\) and \(\alpha < 0.5\), with the former setting the lower threshold in the feasible \(\tilde{a}\). Consider the example where the mass of the boson is \(\mu_S = \mu_V = 10^{-13} \text{eV}\). For PGA-0, the main constraint is at \(M_{\rm BH} \sim 10^2 M_\odot\), which corresponds to \(\tilde{a}_s \simeq 0.3\). However, for PGA-1, the main constraint is \(M_{\rm BH} \sim 10 M_\odot\) (as discussed in the first point), with the corresponding \(\tilde{a}_s\) being about an order of magnitude smaller, $\sim 0.03$. It is obvious that the lower threshold for PGA-1 is smaller compared to PGA-0. 
    \item Third, as shown in Fig.~\ref{paraspace_scalar} and Fig.~\ref{paraspace_vector}, the PGA-1 located in the parameter space $(\mu_V,M_{\rm BH})$ has a significantly wider detectable area, compared to PGA-0 in the parameter space $(\mu_S,M_{\rm BH})$  (taking the same spin distribution). This is due to the more effective GW emission rate of PGA-1 than PGA-0 for the same $\alpha$, which improves the detectability of PGA-1 through a lower peak frequency. It makes SGWB enter the sensitivities of powerful detectors such as LISA, Taiji, and TianQin, as shown by the spectrum for $\mu_V=10^{-12} ~\rm eV$ and $M_{\rm BH}=10~M_{\odot}$ in Fig.~\ref{GW_2a}. 
    \begin{figure}[htbp]
			\centering
			\includegraphics[width=0.9\textwidth]{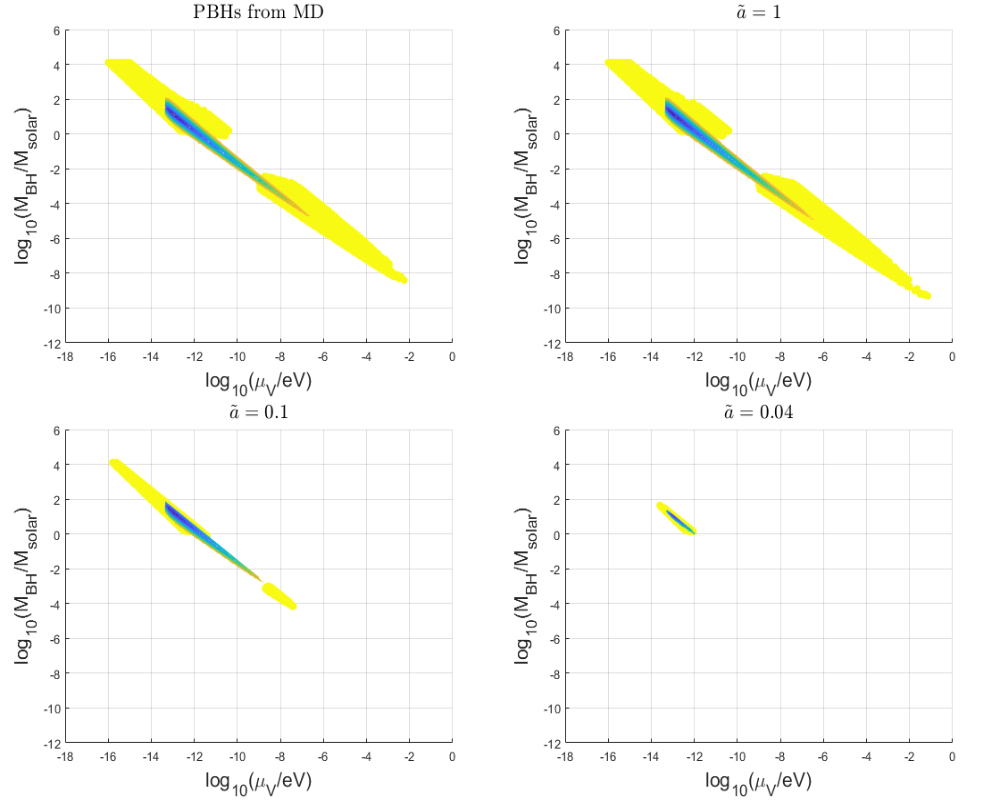} 
			\caption{Constraints (blue and green bands) and prospects (yellow bands) of PGA-1,  taking different spins $\tilde a$. }
    \label{paraspace_vector}
		\end{figure}

    \end{itemize}

Note that the fundamental difference between the two types of PGAs lies in the different \(f_{\rm peak}\) caused by \(\tau_{\rm GW}\).

\section{Discussions and Conclusions}

Gravitational atoms have been an interesting interdisciplinary field between astronomy and particle physics and have inspired a lot of studies. However, if the cores of GAs are PBHs, then cosmology joins. This scenario has not been carefully studied yet, probably due to the lack of evidence of PBHs, particularly PBHs with a large spin. However, observations really call for primordial origins of black holes and moreover PBH relics are well expected in the early universe. 

In this article, our aim is to investigate the general features of GWs from PGAs. To make the discussion more concrete, we consider a representative monochromatic mass scenario in which PBHs form during a short MD era. In this scenario, the PBHs acquire a large spin at formation and do not lose it through accretion. We compute the GW signals generated by the boson clouds and, through approximate analysis, we find that PGAs in the universe leave an imprint as an SGWB. This spectrum is characterized by a rise $\propto f^3$ followed by a decline $\propto f^{-1}$, distinguishing it from other sources of GW signals. Furthermore, we provide constraints on the abundance of PGAs. We find that compared to bare PBHs with $M_{\rm BH} \sim \mathcal{O}(10) M_\odot$,  PGA-0 with \( \mu_S \sim \mathcal{O}(10^{-13}) \text{ eV} \) and PGA-1 with \( \mu_V \sim \mathcal{O}(10^{-12}) \text{ eV} \) are imposed stronger constraints. Additionally, considering future GW detectors such as LISA, Taiji, and TianQin, we identify a narrow and elongated parameter space. In the near future, we will know whether these PGAs are detected or further constrained. 

This is a tentative study, and many things need to be clarified or improved further. The most urgent one, we think, is the effect of non-rotating matter accretion around the PBHs formed during the MD era. Specific simulation (based on the classical field that behaves as matter) indicates that it may reduce the initially large dimensionless spin. Although there is no comprehensive research about the PBH spin evolution during the MD era, we could estimate the effect of accretion. In Ref. \cite{DeLuca:2021pls}, the secondary infall model was used to consider the spherically symmetric accretion of non-spinning PBHs, where the PBH mass grows as $M_{\rm BH} \propto a \propto t^{2/3}$. Assuming that the rotating PBH exhibits a similar accretion behavior, we then have \(\tilde{a} \propto a^{-1}\) according to the conservation of the total angular momentum. Clearly, even if the initial dimensionless spin is at the maximum value of 1, it will be reduced to approximately 0.1 after two e-folds.

In this paper, we focus on a short MD era to prevent spin reduction. However, it is worth noting that our results remain significant when extended to a longer MD era. The constraints we place on PBHs and the prospects we discuss are applicable to those cores formed near the reheating period ($M_{\rm BH}\simeq M_{\rm max}$). These black holes, which might have vital clues about the early universe, have the potential to help us understand the origins of cosmic history.

Moreover, the SGWB energy density frequency spectrum $\Omega_{\rm GW}(f)$ is very sensitive to the core mass, as $\frac{d\tilde{E}}{dt} \propto M_\text{BH}^{14}$ for PGA-0 and $\frac{d\tilde{E}}{dt} \propto M_\text{BH}^{10}$ for PGA-1 as shown in Eq. \eqref{dedt_scalar} and Eq. \eqref{dedt_vector}. This implies that the evolution of core mass could play a crucial role, including its late accretion and cosmic evolution. However, the energy extraction from cores by superradiant clouds does not have a significant impact. This is because the superradiance process affects the black hole's mass by at most $10.78\%$ \cite{hui2208black}, which does not change the order of magnitude of $\Omega_{\rm GW}(f)$.

\section*{Acknowledgements}

TL is supported in part by the National Key Research and Development Program of China Grant No. 2020YFC2201504, by the Projects No. 11875062, No. 11947302, No. 12047503, and No. 12275333 supported by the National Natural Science Foundation of China, by the Key Research Program of the Chinese Academy of Sciences, Grant No. XDPB15, by the Scientific Instrument Developing Project of the Chinese Academy of Sciences, Grant No. YJKYYQ20190049, and by the International Partnership Program of Chinese Academy of Sciences for Grand Challenges, Grant No. 112311KYSB20210012.

\appendix

\section{PBHs Formed during a MD Epoch}\label{MD:PBH}

\subsection{The PBH Mass} \label{Appendix PBH mass}
Consider a perturbation with scale $k$, its density contrast $\delta(k) \equiv (\rho-\bar\rho)/\bar\rho$  describes the fluctuation relative to the background. Its behavior after  entering the Hubble horizon is quite interesting, because causality becomes established. The time of horizon crossing is $t_H(k)$, which is determined by the equality between the physical length $a(t_H(k))/k$ and the Hubble horizon length $1/H(t_H(k))$, therefore $t_H(k)=\frac{8a_m}{3k^3}$ (the subscript $H$ emphasizes that this quantity is defined at the time of horizon crossing, and $a_m$ is a constant). Primordial fluctuations are random and generally follow a Gaussian distribution. At $t=t_H(k)$, the probability distribution of the density contrast is 
\begin{align}
\label{eq_pdelta}
    P(\delta_H)=\frac{1}{\sqrt{2\pi}\sigma_H}
    \exp \left( 
    -\frac{\delta_H^2}{2\sigma_H(k)^2}
    \right),
\end{align}
where $\sigma_H$ is much smaller than 1 in general, thus the majority of perturbations satisfies $\delta_H \sim \sigma_H \ll 1$.
After the perturbation crosses the Hubble horizon, it continues to expand following the Hubble flow, but the expansion gradually slows down due to the gravity. Consequently, the density contrast increases, a process that can be described using the results of linear perturbation theory as $\delta =\delta_H a(t)/a(t_H)\propto t^{2/3}$. When the perturbation has expanded to its maximum size (at $t=t_d$), it decouples from the background, and then the perturbation collapses. Since $\delta \sim 1$ typically occurs at $t=t_d$ (which implies $t_d(k)\simeq t_H(k)\delta_H(k)^{-3/2}=\frac{8a_m}{3k^3}\delta_H(k)^{-3/2}$), this collapse cannot be described by linear theory. The collapse completes at $t=t_c$, with $t_c$ generally being of the same order as $t_d$ \footnote{The details have been discussed in Ref.~\cite{gunn1972infall} for isotropic collapse and Ref.~\cite{harada2016primordial} or anisotropic collapse. } . 

The probability $\beta(k)$ for collapsing to a PBH is discussed in Appendix \ref{MD:prsd}. It actually gives the corresponding energy fraction of the formed PBH if the resulting PBH mass $M_{\rm BH}$ is the Hubble mass at horizon crossing $M_H(k)=\frac{M_{\rm PL}^2}{2H(t_H(k))} \propto t_H(k)$. Therefore, the PBH born later has a larger mass. 

For the perturbation with $t_c(k)>t_{\rm end}$, the universe enters the RD era before the collapse completes. This significantly reduces the probability of formation of PBH. Therefore, we only consider the perturbations satisfying
\begin{align}
t_c(k) \simeq t_d(k) \simeq t_H(k) \delta_H(k)^{-3/2}\in (t_{\rm ini},t_{\rm end}).
\end{align}
The PBH mass spectrum starts from $M_{\rm min}= M_H(t_{\rm ini})$, since the scale $k_{\rm ini}$ entering horizon at $t_{\rm ini}$ is supposed to have sufficient time to evolve and collapse in the MD era. 
The PBH mass spectrum ends at $M_{\rm max}= M_H(t_{\rm max})$, which originates from perturbations with $k = k_{\text{max}}$, defined by $t_c(k_{\rm max})=t_{\rm end}$.
Then, the corresponding PBH mass is given by 
\begin{align}\label{max:mass}
 M_\text{max}=M_H(t_\text{end})
 \frac{t_H(k_{\rm max})}{t_\text{end}}
 \simeq M_H(t_\text{end})\delta^{3/2}_H(k_\text{max})
 \simeq M_H(t_\text{end})\sigma^{3/2}_{\text{max}},
\end{align} 
where $\sigma_{\rm max}\equiv \sigma_H(k_{\rm max})$.

\subsection{The Production Rate and Spin Distribution}\label{MD:prsd} 

Harada et al. in Ref.~\cite{harada2016primordial} considered that a uniformly distributed spherically symmetric region that, upon entering the horizon, undergoes anisotropic collapse, as shown in
\[
\begin{cases}
    r_1= (a-\alpha b)q_1 \\
    r_2= (a-\beta b)q_2~,\\
    r_3= (a-\gamma b)q_3 
\end{cases}
\]
where $\alpha,\beta,\gamma$ are constant satisfying  $\alpha>\beta>\gamma$ and $\delta_H=\alpha+\beta+\gamma >0$ . The probability distribution function for them is given by Doroshkevich \cite{harada2016primordial} as
\begin{align}
    w(\alpha,\beta,\gamma)=-\frac{27}{8\sqrt{5}\pi\sigma_3^6}
    \exp{\left[
        -\frac{3}{5\sigma_3^2} (\alpha^2+\beta^2+\gamma^2) - \frac{1}{2}(\alpha\beta+\beta\gamma+\gamma\alpha)
    \right]} (\alpha-\beta)(\beta-\gamma)(\gamma-\alpha).
\end{align}
It is consistent with Eq. (\ref{eq_pdelta}) and $\sigma_H = \sqrt{5}\sigma_3$.
Due to the anisotropy of collapse, this overdense region forms a pancake. According to the hoop conjecture, if it satisfies $h(\alpha,\beta,\gamma)\lesssim1$, it eventually collapses into a black hole, where
\begin{align}
    h(\alpha,\beta,\gamma) \equiv \frac{2}{\pi} \frac{\alpha-\gamma}{\alpha^2}
    E \left( 
    \sqrt{1- \left( \frac {\alpha-\beta}{\alpha-\gamma} \right)^2} 
    \right),
\end{align}
and $E(e)$ is the complete elliptic integral of the second kind. The production rate $\beta_\text{ani}$ is given by
\begin{align}
    \beta_\text{ani} = \int^\infty_0 d\alpha \int^\alpha_{-\infty} d\beta \int^\beta_{-\infty} d\gamma \theta [\delta_H(1-h(\alpha,\beta,\gamma)] \omega(\alpha,\beta,\gamma),
\end{align}
which is best fit by $\beta_\text{ani}=0.056\sigma_H^5$ for small $\sigma_H$.

Subsequently, the effect of spin on black hole collapse was considered in Ref. \cite{harada2017spins}. 
According to their assumptions, the spin of the product after the collapse of the overdense region ultimately receives two contributions
    \begin{align}
        \tilde{a} = \tilde{a}_{(1)} + \tilde{a}_{(2)} \approx \text{max} (\tilde{a}_{(1)},\tilde{a}_{(2)})~,~\,
    \end{align}
where
    \begin{gather} 
		\label{pbhspin}
		\tilde{a}_{(1)} = \frac{2}{5} \sqrt{\frac{3}{5}} q (\delta_H)^{-1/2}\ \text{and}\
		\tilde{a}_{(2)} = \frac{2}{5} I (\delta_H)^{-3/2}.
    \end{gather}
$q$ is a dimensionless constant describing the initial shape of the perturbation ($q=0$ refers to a sphere), and its distribution is unknown to our knowledge, which brings theoretical uncertainty to the quantitative study of the PBH spin distribution. $I$ is another dimensionless constant related to the perturbation power spectrum. According to the calculations of several power spectrum in \cite{peebles1969origin}, it can be assumed that $I\sim 1$.  In this paper, we always take $I=1$. Black holes must satisfy $\tilde{a}<1$, therefore, they argued that the pancake satisfying $\tilde{a}<1$ will eventually form a black hole, which imposes a threshold for PBH formation, $\delta_\text{th}=\text{max}(\delta_\text{th}^{(1)} , \delta_\text{th}^{(1)}) \equiv \text{max}(\frac{12}{125}q^2 , (\frac{2}{5} I \sigma_H)^{\frac{2}{3}})$ from Eq.~(\ref{pbhspin}) by setting $\tilde a=1$.
It is seen that, $\delta_\text{th}=\delta_\text{th}^{(1)}$ when $q>q_c \equiv \sqrt{\frac{2}{3}}(\frac{5}{2})^{7/6}I^{1/3}\sigma_H^{1/3} \sim \sigma_H^{1/3}$; $\delta_\text{th}=\delta_\text{th}^{(2)}$ when $q<q_c$. Combining the effects of spin and anisotropic collapse on the PBH formation, the probability for a perturbation collapsing to a PBH can be calculated by 
\begin{align}
    \beta_\text{ani+spin} \simeq \int^\infty_0 d\alpha \int^\alpha_{-\infty} d\beta \int^\beta_{-\infty} d\gamma \theta [\delta_H(\alpha,\beta,\gamma)-\delta_{th}] \theta [\delta_H(1-h(\alpha,\beta,\gamma)] \omega(\alpha,\beta,\gamma)~.~\,
\end{align}
The numerical results are shown in Figure 5 in Ref. \cite{harada2017spins}, and  the semi-analytical results are provided below.
For $q>q_c (\delta_{th}=\delta_{th}^{(1)})$, we have
\begin{equation}
    \beta_\text{(1)ani+spin} \simeq
    \begin{cases}
        3.244 \times 10^{-14} \cfrac{q^{18}}{\sigma_H^4} \exp \left[ -0.004608\cfrac{q^4}{\sigma_H^2} \right] & \quad (\sigma_H \lesssim 0.04) \\
        0.056\sigma_H^5 & \quad (0.04 \lesssim \sigma_H \lesssim 0.2)~.~\,
    \end{cases}
\end{equation}
While for $q<q_c (\delta_{th}=\delta_{th}^{(2)})$, we have
\begin{equation}
    \beta_\text{(2)ani+spin} \simeq
    \begin{cases}
        1.921 \times 10^{-7} I^6 \sigma_H^2 \exp \left[ -0.1474\cfrac{I^{4/3}}{\sigma_H^{2/3}} \right] & \quad (\sigma_H \lesssim 0.005) \\
        0.056\sigma_H^5 & \quad (0.005 \lesssim \sigma_H \lesssim 0.2)~.
    \end{cases}
\end{equation}

According to Eq.~(\ref{sigma_max}), we can estimate $\sigma_H$. Noting that we are only concerned with the upper limit of $f_\text{PBH}$, which lies between $10^{-10}$ and 1, as shown in Fig. \ref{newconstraint} and Fig. \ref{newconstraint_vector}, we estimate $0.002\lesssim \sigma_H \lesssim 0.08$. Of course, when $10^{-2.5}M_\odot \lesssim M_{\rm BH} \lesssim 10^{-0.5}M_\odot$,  $f_\text{PBH}$ can be much less than $10^{-10}$. However, we find that this range does not yield a detectable SGWB and thus does not affect our discussion.

For $0.002\lesssim \sigma_H \lesssim 0.04$, it is not difficult to observe $\beta_\text{(2)ani+spin} \gg \beta_\text{(1)ani+spin}$, which implies that most PBHs originate from overdense regions that satisfy $q<q_c$. Taking into account statistical averaging, the PBH production rate is approximately equal to $f(q_c)\beta_\text{ani+spin(2)} \simeq f(q_c) \beta_\text{ani}$ (with a difference of approximately an ${\cal O}(1)$ factor when $0.002 \lesssim \sigma_H \lesssim 0.005$). The factor $f(q_c)$ represents the proportion of $q < q_c$ and depends on the initial distribution of $q$, which is unknown to our knowledge.
For $0.04 \lesssim \sigma_H \lesssim 0.08$, $\beta_\text{ani+spin} = \beta_\text{ani}$, and the production rates for $q<q_c$ and $q>q_c$ are equal. 
Note that the range of $q_c$ in which we are interested satisfies $ 0.3 \lesssim q_c \lesssim 1$ by estimation. As a preliminary exploration, we assume that the vast majority of the overdense regions satisfy $q < q_c$, thus $f(q_c) = 1$. Consequently, in the scenario we consider, $\beta_\text{ani+spin}=\beta_\text{ani}$, indicating that PBH formation is independent of spin.

\begin{figure}[htbp]
			\centering
			\includegraphics[width=0.7\textwidth]{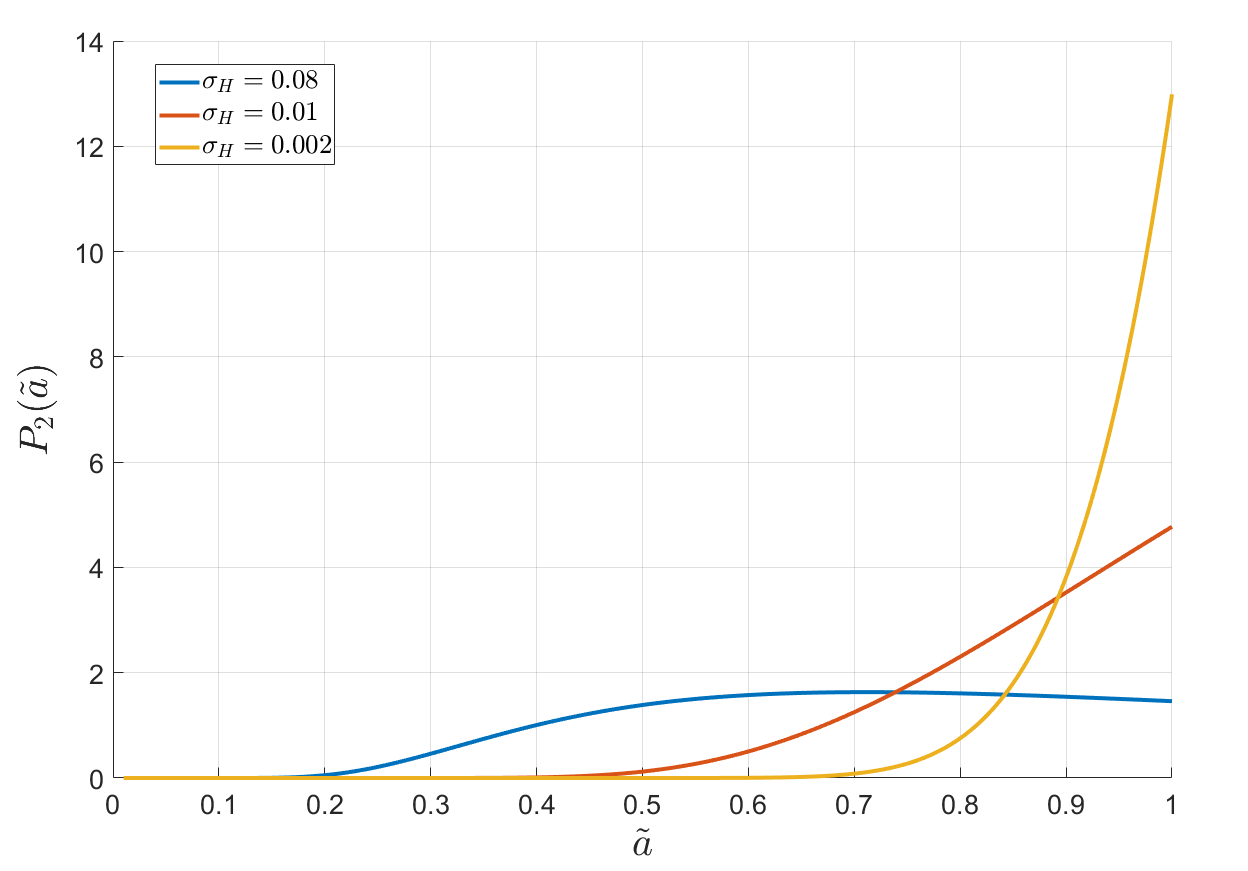} 
			\caption{The spin distribution of PBHs. The lines with different color denote the different $\sigma_H$.}
                \label{P2a}
		\end{figure}

In Ref.~\cite{kokubu2018effect}, the PBH production probability of spherically symmetric but inhomogeneous distributed overdense regions is calculated, being fit by $\beta_\text{inhom} \simeq 3.6979 \sigma_H^{3/2}/\Sigma$, where $\Sigma \sim {\cal O}(1)$.
This fits well when $\sigma_H \lesssim 0.05$, and there is a difference of approximately an ${\cal O}(1)$ factor when $0.05 \lesssim \sigma_H \lesssim 0.08$. Therefore, $\beta_\text{inhom} \simeq 3.6979 \sigma_H^{3/2}$.
To combine the effects of the inhomogeneity and anisotropy on the PBH formation (the spin has no effect), we further assume that all overdense regions are very close to sphericity, so that $q \ll q_c \sim 1$. In this case, the prodution rate is expected as the simple multiplication of $\beta_\text{inhom}$ and $\beta_\text{ani}$,
\begin{align}
    \beta_\text{ani+spin+inhom} = \beta_\text{inhom} \times \beta_\text{ani}
    \simeq 0.2055 \sigma_H^{13/2}.
\end{align}
At the same time, under the assumption that $q \ll q_c$, for the majority of PBHs, $\tilde a_{(2)} >\tilde a_{(1)}$ (see Fig. 2 in Ref.~\cite{harada2017spins}). Thus, $\tilde{a}\simeq \tilde a_{(2)}$. According to the results in Ref.~\cite{harada2017spins}, we can infer that the spin distribution of PBHs at formation is
 \begin{gather}
    P_2(\tilde{a}) \propto \frac{1}{\tilde{a}^{5/3}} \exp 
		\left(
			- \frac{1}{2 \sigma_H^{2/3}} \left( \frac{2}{5} I\right)^{4/3} \frac{1}{\tilde{a}^{4/3}}
		\right)~.
\end{gather}
We have plotted $P_2(\tilde{a})$ for $\sigma_H=0.08$, $\sigma_H=0.01$ and $\sigma_H=0.002$, as shown in Fig. \ref{P2a}. It's evident that the smaller the $\sigma_H$, the more concentrated the initial spins of PBHs are around $\tilde{a}=1$. This is because a smaller perturbation requires more time to reach the maximum expansion, giving the dimensionless spin ample time to grow (see Ref. \cite{harada2017spins} for details).

\bibliographystyle{unsrt}  
\bibliography{ref} 
\end{document}